\documentclass[twocolappendix,numberedappendix,apj]{emulateapj}
\usepackage{amsmath,amssymb,amsfonts,bm}
\usepackage{epstopdf}
\usepackage{epsfig}
\usepackage{natbib,caption2}
\usepackage{graphicx}   
\usepackage{float}
\usepackage{longtable}
\usepackage{graphics}
\usepackage{hyperref}
\usepackage{color}
\usepackage{calc}
\usepackage{threeparttable}

\graphicspath{{figures/}}

\newcommand \beq{\begin{equation}}
\newcommand \eeq{\end{equation}}
\newcommand \bey{\begin{eqnarray}}
\newcommand \eey{\end{eqnarray}}
\newcommand \Myr{\, {\rm Myr} }
\newcommand \Gyr{\, {\rm Gyr} }

\newcommand \pc{\, {\rm pc} }
\newcommand \kpc{\, {\rm kpc} }

\newcommand \msun{M_\odot}
\newcommand \Msun{M_\odot}
\newcommand \rsun{R_\odot}

\newcommand \kms{\, {\rm km \, s}^{-1} }

\newcommand \vzero{{\bf v}_{\rm 0}}

\begin{document}

\title{Ring galaxies through off-center minor collisions by tuning bulge-to-disk mass ratio of progenitors}

\shorttitle{Off-center collisional ring galaxies}
\shortauthors{Chen et al.}

\author{Guangwen Chen\altaffilmark{1,2}, Xufen Wu\altaffilmark{1,2}, Xu Kong\altaffilmark{1,2}, Wen-Juan Liu\altaffilmark{3,4}, HongSheng Zhao\altaffilmark{5}}
\thanks{guangwen@mail.ustc.edu.cn, xufenwu@ustc.edu.cn}
\altaffiltext{1}{CAS Key Laboratory for Research in Galaxies and Cosmology, Department of Astronomy, University of Science and Technology of China, Hefei 230026, China; }
\altaffiltext{2}{School of Astronomy and Space Science, University of Science and Technology of China, Hefei 230026, China}
\altaffiltext{3}{Yunnan Observatories, Chinese Academy of Sciences, Kunming, Yunnan 650011, China}
\altaffiltext{4}{Key Laboratory for the Structure and Evolution of Celestial Objects, Chinese Academy of Sciences, Kunming, Yunnan, 650011, China}
\altaffiltext{5}{School of Physics and Astronomy, University of St Andrews, North Haugh, Fife, KY16 9SS, UK}

\begin{abstract}
  Collisional ring galaxies (CRGs) are formed through off-center collisions between a target galaxy and an intruder dwarf galaxy. We study the mass distribution and kinematics of the CRGs by tuning the bulge-to-disk mass ratio ($B/D$) for the progenitor; i.e., the target galaxy. We find that the lifetime of the ring correlates with the initial impact velocity vertical to the disk plane (i.e., $v_{z0}$). Three orbits for the collisional galaxy pair, on which clear and asymmetric rings form after collisions, are selected to perform the \textit{N}-body simulations at different values of $B/D$ for the progenitor. It is found that the ring structures are the strongest for the CRGs with small values of $B/D$. The S\'{e}rsic index, $n$, of the central remnant in the target galaxy becomes larger after collision. Moreover, the S\'{e}rsic index of a central remnant strongly correlates with the initial value of $B/D$ for the progenitor. A bulge-less progenitor results in a late-type object in the center of the ring galaxy, whereas a bulge-dominated progenitor leads to an early-type central remnant. Progenitors with $B/D\in [0.1,~0.3]$ (i.e., minor bulges) leave central remnants with $n\approx 4$. These results provide a possible explanation for the formation of a recently observed CRG with an early-type central nucleus, SDSS J1634+2049. In addition, we find that the radial and azimuthal velocity profiles for a ring galaxy are more sensitive to the $B/D$ than the initial relative velocity of the progenitor.

\end{abstract}


\section{Introduction}

Ring galaxies are peculiar galaxies with a bright ring structure that is composed of gas and stars, in which the diameters of the rings could reach up to a few tens to a hundred $\kpc$ \citep{ghosh2008ugc}. The first ring galaxy observed by \citet{zwicky1941cluster} is the now famous ``Cartwheel'' galaxy, close to which there are three companion galaxies. More ring galaxies have been discovered in the decades that followed (e.g., \citealt{vorontsov1959atlas, arp1966atlas, de1975southern, theys1976ring, conn2011new}). \citet{theys1976ring} classified the ring galaxies into three sub-classes: (i) empty ring galaxies without a central nucleus (RE), (ii) ring galaxies with an off-center nucleus (RN), and (iii) knotty ring galaxies (RK). \citet{few1986ring} made another classification for a larger sample of $69$ ring galaxies according to the morphology of the rings. Galaxies with smooth symmetric rings are of O-type, while galaxies with an off-center nucleus and a knotty ring fall into P-type. Polar ring galaxies (PRGs) were discovered later \citep{schechter1978ngc,whitmore1990new,afanasiev2011scorpio}. Most of the PRGs are gas-poor early-type (E/S0) galaxies, with a gas-rich ring or disk structures at a large angle to the major axes of the galaxies.  

Several theories have been developed to explain the formation of ring galaxies, including Lindblad resonances driven by the galactic bars \citep{buta1996galactic, buta1999structure}, accretion from nearby gas-rich galaxies \citep{schweizer1983colliding}, merger of galaxies \citep{bekki1998formation} and encounters between a disk galaxy and a companion dwarf galaxy \citep{lynds1976interpretation,theys1976ring}. The resonance theory explains the formation of O-type inner and outer galactic rings that do not have nearby companion galaxies. Many of the polar ring galaxies can be represented by cold accretion \citep{reshetnikov1997global,bournaud2003formation,maccio2006radial} or the galaxies merging \citep{bekki1998formation}. The P-type galaxies with a companion galaxy are collisional ring galaxies \citep[hereafter CRGs,][]{few1986ring,hernquist1993spokes,appleton1996collisional,horellou2001model}.

The parameter space of the CRGs have already been extensively investigated by means of numerical simulations. The relevant parameters include initial relative velocity \citep{fiacconi2012adaptive}, impact parameter $b$ (the minimum distance between the target galaxy and the intruder galaxy, \citealt{toomre1978interacting, fiacconi2012adaptive}), the inclination angle \citep{lynds1976interpretation, ghosh2008ugc}, the mass ratio between the intruder galaxy and the target galaxy \citep{hernquist1993spokes, horellou2001model}, and the gas fraction and star formation in the target disk \citep{mapelli2012ring}.

The two main stellar components of galaxies are disks and bulges . The distribution of bulge-to-total stellar-mass ratio, $B/T$, is tightly related to the formation of galaxies and the mass assembly histories. The recent work of \citet{smith2012numerical} considered a model with $B/T \approx 1/3$ to reproduce the thick bridge structure for Auriga's Wheel. However, thus far, a systematic study of the impact of the variation of $B/T$ on the formation of the CRGs is still lacking. In this paper, we present an extensive analysis of the formation of CRGs by considering target galaxy models with different bulge-to-disk mass ratios, $B/D$, which is related to the $B/T$ ratio by $B/T=B/(B+D)$. 

Our models of the target and the companion galaxies are constructed in equilibrium in Sec. \ref{sec-models}. To avoid any disturbance of ring formation from intrinsic instability of the parent galaxy (e.g., the inner or outer Lindblad resonances), the models are freely evolved for long enough (i.e., $2~\Gyr$) to test the stability of the models and to obtain stable final products. Using the stable galaxy models, we launch the intruder galaxy from different orbits to encounter with the target galaxy in Sec. \ref{sec-parameters}. We study the morphology and kinematics of the ring galaxy with different values of $B/D$ of the parent galaxy in Sec. \ref{sec-bd}. The value of $B/D$ varies from $0.0$ to $1.0$ with an interval of $0.1$, which represents pure-disk galaxies to bulge-dominated galaxies. Finally, we discuss an application of our models to a recently observed AGN-host CRG, SDSS J1634+2049 \citep{liu2016sdss} and conclude in Sec. \ref{conclusion}.

\section{Models}\label{sec-models}
The first successful modeling of ring galaxies formed through galaxy encounter was implemented by \citet{lynds1976interpretation}, who supposed that a companion dwarf galaxy collides with the center of a disk galaxy in the direction more or less perpendicular to the disk plane. The stars in the disk are attracted to the center when the dwarf galaxy is approaching. These stars leave the galactic center after the companion goes through the disk and the orbits of the stars have been changed permanently. Accordingly, a ring-like density structure spreads outwards from the galactic center. The dynamic age of the ring galaxies is about a couple of hundred $\Myr$. \citet{theys1977ring} and \citet{toomre1978interacting} found that the impact parameter sensitively affects the morphology of the ring. 

A series of numerical simulations have shown that the offset of a nucleus and a lopsided ring can be generated through an off-center collision (e.g., \citealt{lynds1976interpretation,theys1976ring,appleton1996collisional} for a review, \citealt{mapelli2012ring} and \citealt{smith2012numerical}). Moreover, \citet{appleton1987models} revealed that a small and loosely bound intruder galaxy could be disrupted after a collision. In the mushroom-shaped galaxies (e.g., the Sacred Mushroom; \citealt{arp1987catalogue}, a companion galaxy has been disrupted by the collision and has formed a stellar stream \citep{wallin1994observations}. Meanwhile, head-on collisions with different mass ratios of intruder-to-target galaxies have been studied by \citet{gerber1996stellar}. The density of the ring is larger and the propagation velocity of the ring is higher for a more massive intruder galaxy.

Using a model with a nearly central collision at a non-zero inclination angle between the most distant companion, G3, and the target host galaxy, \citet{horellou2001model} successfully reproduced the Cartwheel ring galaxy. The collisional system has been further evolved for a few hundred $\Myr$ after the formation of galactic ring in \citet{mapelli2008ring}. A ring structure forms around $100$--$200\Myr$ after a collision and keeps expanding afterwards, and therefore produces a very extended galactic disk that can be observed as a giant low surface brightness galaxy (GLSB). \citet{mapelli2008ring} found that the RE galaxies can be generated through off-center collisions, even for small inclination angles. The impact parameter plays an important role in the formation of a ring or an arc. Recently, a newly observed ring galaxy, Auriga's Wheel, has been successfully modeled by \citet{smith2012numerical}, with a relatively low initial velocity of $\approx 150\kms$ to allow for the formation of the observed bridge. In addition, the stellar bridge mainly originates from the disk and bulge of the target galaxy rather than from the companion.

Recently, a new ring galaxy has been studied by \citet{liu2016sdss}. This is a host galaxy of an active galactic nucleus (AGN), SDSS J163459.82+204936.0. A ring-like structure (i.e., an arc structure surrounding the central nucleus) has been found at a projected distance of around $16~\kpc$ to the galactic center. In the extended direction of the galactic center to the arc, there are two dwarf galaxies, namely C1 and C2, at a projected galactocentric distance of around $35.7~\kpc$ and $44.0~\kpc$, respectively. The redshifts of the two dwarf galaxies are close to that of the host galaxy of SDSS J1634+2049, thus they are considered as companion galaxies of the AGN-host galaxy. The arc structure is proposed to be the dense region of a knotty ring around the AGN. The host galaxy of SDSS J1634+2049 seems to be a CRG that formed in the collision with C1 galaxy. The C2 galaxy is too faint. Consequently, it is not massive enough to generate a collisional ring in the host galaxy.

To conveniently apply the numerical results to the newly observed CRG system, we shall perform a series of numerical simulations that are based on the basic parameters of the host galaxy of SDSS J1634+2049 and the companion C1 galaxy, including the masses and the relative projected distance of the two galaxies. However, our numerical simulations are quite generic and can be used to study the formation of other CRGs.

\subsection{Galaxy Models and Numerical Tools}\label{ics}
\subsubsection{The Target Galaxy}
The target galaxy model in our simulations is a massive disk galaxy that contains a thin stellar disk, a stellar bulge and a dark matter (DM) halo. The initial conditions (ICs) of the target galaxy are generated by using the \textit{BUILDGAL} code, as described in \citet{hernquist1993n}, which generates self-consistent \textit{N}-body disk-bulge-halo galaxy models. \textit{BUILDGAL} has been frequently used to build equilibrium multicomponent disk galaxies \citep{boily2001efficient,penarrubia2010progenitor}.

We consider the density profile of the bulge component introduced by \citet{hernquist1990analytical},
\begin{equation}
 \rho_{\rm{b}}(m)=\frac{M_{\rm{b}}}{2 \pi ac^2} \frac{1}{m(1+m)^3},
 \label {eq2-1}
\end{equation}
where $M_{\rm{b}}$ is the bulge mass, $a$ is the scale length along the major axis, $c$ is the scale length along the minor axis, and
\begin{equation}
m^2=\frac{x^2+y^2}{a^2}+\frac{z^2}{c^2},
 \label {eq2-2}
\end{equation}
where the ratio of scale lengths on $x,y,z$ axis is $1:1:0.5$ (see Table \ref{galmodels}) and it is an oblate system. The intrinsic shapes of a sample of 83 CALIFA bulges have been studied by \citet{costantin2017intrinsic}, and $66\%$ of them are oblate systems. Therefore, the axial ratio for the target model galaxy that is used here is reasonable. 

The density profile of the stellar disk follows axisymmetric exponential distribution and is given by: 
\begin{equation}
 \rho_{\rm{d}}(R,z)=\frac{M_{\rm{d}}}{4 \pi h^2 z_0} \exp(-R/h) {\rm sech}^2(z/z_0),
 \label {eq2-3}
\end{equation}
where $M_{\rm{d}}$ is the disk mass, and $h$ and $z_0$ are the radial and vertical scale length of the disk, respectively. A Toomre parameter, $Q=1.5$ is chosen at the radius of the solar neighborhood, $\rsun=8.5\kpc$, to normalize the radial velocity dispersion. In the model, $Q \ge 1.5$ in all radii to ensure that the disk is stable \citep{toomre1964gravitational}. Before testing the effects of different values of $B/D$, we shall explore the parameter space, including the initial positions and relative velocity of the galaxy pair, impact parameter and the concentration of the target galaxy. $B/D$ is fixed at a value of $0.25$ and the orbits that can generate a CRG are selected in this stage. A model with $B/D=0.25$ represents a disk-dominated galaxy with a moderate bulge component. The total stellar mass of the target galaxy, $M_{\rm{*,host}}$, is $1.34\times 10^{11} \msun$, which presents a massive disk galaxy such as M31 and the host galaxy of SDSS J1634+2049. The masses of bulge and disk particles are $1.34\times10^4\msun$ and $2.68\times 10^4\msun$, respectively. 

The density profile of a dark matter halo is a truncated isothermal sphere, 
\begin{equation}
 \rho_{\rm{h}}(r)= \frac{M_{\rm h}}{2 \pi^{3/2}} \frac{\alpha}{r_{\rm t}} \frac{\exp(-r^2 / r^2_t)}{r^2+r_c^2},
 \label {eq2-4}
\end{equation}
where $M_{\rm{h}}$ is the halo mass, $r_{\rm{t}}$ and $r_{\rm{c}}$ are the tidal radius and the core radius of the halo, respectively. The normalization constant $\alpha$ is defined as follows: 
\begin{equation}
 \alpha = \left \{ 1-\sqrt{\pi}q \exp(q^2)[1-{\rm erf}(q)]\right \}^{-1},
 \label {eq2-5}
\end{equation}
where $q = r_{\rm{c}} / r_{\rm{t}}$ and ${\rm erf}(q)$ is the error function. The parameters of the disk model are listed in the lines $2$--$7$ of Table \ref{galmodels}. The total mass of dark halo is $20$ times of $M_{\rm *,host}$, and the overall mass of the host galaxy including stellar and DM particles, $M_{\rm{total,host}}$, is $2.81\times 10^{12} \msun$. The softening for generating the ICs of target galaxy is $3.5\times 10^{-3} \kpc$.

\begin{table}
\centering
\caption[]{Model Parameters of the Target Galaxy ($3^{rd}-8^{th} $ Lines) and the Intruder Galaxy ($9^{th}$ Line)} \vskip 0.2cm
\label{galmodels}
\begin{threeparttable}
\begin{tabular}{lcccccccc}
 \hline
 \hline
  Model & Mass  &  Scale Lengths  & $N_{\rm p}$ & $m$ \\
  & $(10^{10}\msun)$ &$(\kpc)$ &$(10^5)$ &$(10^5\msun)$\\
  \hline
  Bulge &  $M_{\rm{b}}:~ 2.68$  & $a:~ 0.63$  & $20$ & 0.134 \\
  	& &$c:~ 0.315$ & \\
  Disk &$M_{\rm{d}}:~  10.72$& $h:~ 3.50$ & $40$  &0.268\\
        &  &$z_0:~0.70$ &  \\
  Halo &$M_{\rm{h}}:~268.0$&$r_{\rm{c}}:~7.0$  & $2$&134\\
        &  &$r_{\rm{t}}:~ 70.0$ &  \\
  \hline
  Intruder &$M_{\rm P}:~30.0$  &$r_{\rm P}:~1.9$  &$1$ &30\\
    \hline
 \end{tabular}
\begin{tablenotes}
\item 
\begin{flushleft}
\textbf{Note.} The masses are in units of $10^{10}\msun$ and scale lengths are in units of $\kpc$. $N_{\rm p}$ is the number of particles for each component/galaxy in unit of $10^5$ particles and $m$ is the mass of particles in unit of $10^5\msun$.
\end{flushleft}
\end{tablenotes}
\end{threeparttable}
\end{table}

\subsubsection{The Intruder Galaxy}
For simplicity, a Plummer sphere is chosen for simplicity to describe the density profile of the intruder galaxy containing a stellar component and a dark matter halo (\citealt{binney1987galactic}; \citealt{read2006importance}; \citealt{wu2015formation}),
\beq
\rho_{\rm plummer}(r)=\frac{3M_{\rm P}}{4\pi r_{\rm P}^3}(1+\frac{r^2}{r_{\rm P}^2})^{-5/2},
\eeq
where $M_{\rm P}$ is the total mass of the companion galaxy including stellar and DM components, and $r_{\rm P}$ is the scale length of Plummer sphere. The parameters of the companion galaxy are listed in line $8$ of Table \ref{galmodels}. The equal-mass and isotropic \textit{N}-body ICs of the companion galaxy is generated by the \textit{McLuster} code \citep{kupper2011mass}, which has been well examined and widely used in various recent studies \citep{geller2013consequences, pancino2013globular, weidner2013m,  alessandrini2016cosmic} on the dynamics of stellar systems. The mass ratio between the intruder and the target galaxies is $\approx 0.1$, which is the minimal mass ratio to form a collisional ring as suggested by \citet{hernquist1993spokes}. 

\subsubsection{Stability Test of the ICs}\label{stability}
\begin{figure}
\includegraphics[width=85mm]{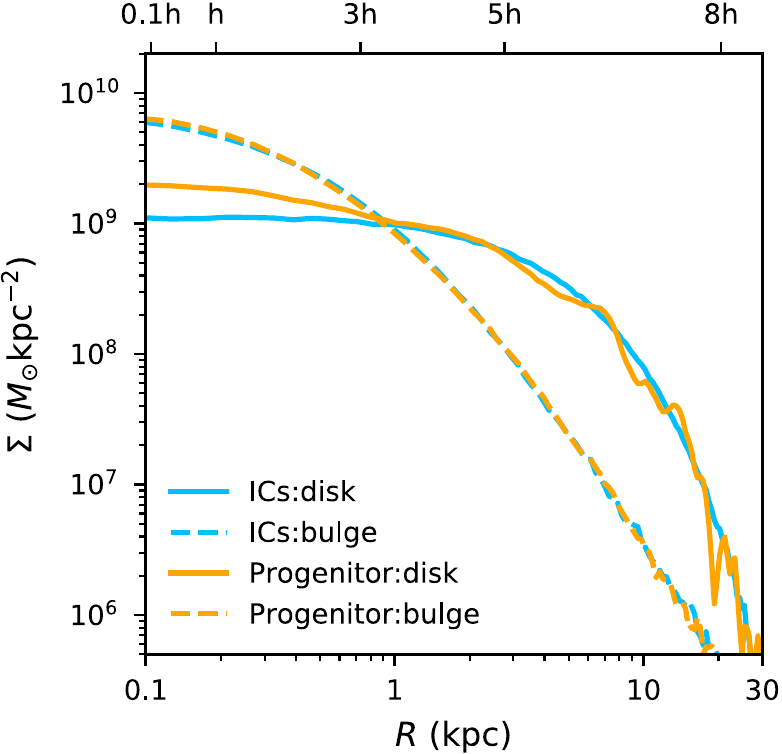}
\caption{Projected mass density profiles of the disk (solid curves) and the bulge (dashed curves) of the target galaxy. The cyan and orange colors represent ICs and model freely evolved for $2~\Gyr$, respectively. }\label{den2d}
\end{figure} 
In general, the cold thin disks are very unstable in general \citep{binney1987galactic}. Many early studies have shown that a pure self-gravitating disk is dynamically unstable and forms a bar on a dynamic timescale \citep{miller1970numerical, hohl1971numerical}. \citet{ostriker1973numerical} found that flattened disk galaxies can become stable if there is a massive and dynamically hot component. Today, there is no doubt that disk galaxies remain stable by adding concentrate centers of dark matter halos \citep{toomre1981amplifies, sellwood2001stability}. In addition, a compact bulge can reduce the buckling instability \citep{sotnikova2005bending}.

The Toomre's parameter is $Q\ge 1.5$ in all radii for the target galaxy, which is near the instability threshold. To avoid any possible influence of the instability on the ring formation, we freely evolve our galaxy models for $2~\Gyr$ in an adaptive mesh refinement (AMR) \textit{N}-body code, \textit{RAMSES} \citep{teyssier2002cosmological}. The collisional simulations are also performed using \textit{RAMSES}. Given that the sizes of our galaxy models are $\approx 100~\kpc$, a box with maximal length of $2500~\kpc$ for a collisional simulation is large enough to include all of the particles that are bound to the system. The following spatial resolutions are chosen for all of the simulations: a minimal resolution of $2500\kpc / 2^7 = 19.5\kpc$ and a maximal resolution of $2500\kpc/ 2^{25}= 0.075 \pc$. The maximal resolution is very similar to that in \citet{renaud2013sub}, where a sub-parsec resolution simulation for the Milky Way is studied. Note that the maximal resolution is not the actual spatial resolution in our simulations. When one is using \textit{RAMSES}, the grid segmentation at a time step either stops when there is only one particle in a cell or when there is more than one particle in a cell but the level of refinement reaches the allowed maximal resolution. In our simulations, the actual maximal level of refinement is $16-19$ (for the models in this section and for the models in Section \ref{sec-bd}), which corresponds to a spatial resolution of $38\pc$ to $5 \pc$. Given that the minimal particle mass is $1.34\times 10^4\msun$ and the central 3-dimensional mass densities of our disk galaxy models are $\approx 10^{10} \msun\kpc^{-3}$, the minimal distance between two disk particles in the densest region of a galaxy is around $(1.34\times 10^4/10^{10})^{1/3}\approx 10 \pc$. This minimal distance agrees with the actual spatial resolution in the simulations. Consequently, the resolution in our simulations is much lower than that in \citet{renaud2013sub}, and our simulations do not suffer from the possible numerical effects caused by very high spatial resolution relative to a comparable high mass resolution. The units are scaled to $M_{\rm simu}=10^{10}~\msun$, $r_{\rm simu}=1.0~\kpc$. The gravitational constant, $G$, is scaled to $G=1$ in the simulations. The crossing time of a particle at radius $5h$ (which encloses over $90\%$ of the baryonic matter) is $t_{\rm cross}\equiv \frac{5h}{v_{\rm circ}(5h)} \approx 45\Myr$, the time scale for free evolution is long enough for the growth of any instability. Here, $v_{\rm circ}(5h)$ is the circular velocity at the radius of $5h$.

\begin{figure*}
\centering  
\includegraphics[width=140mm]{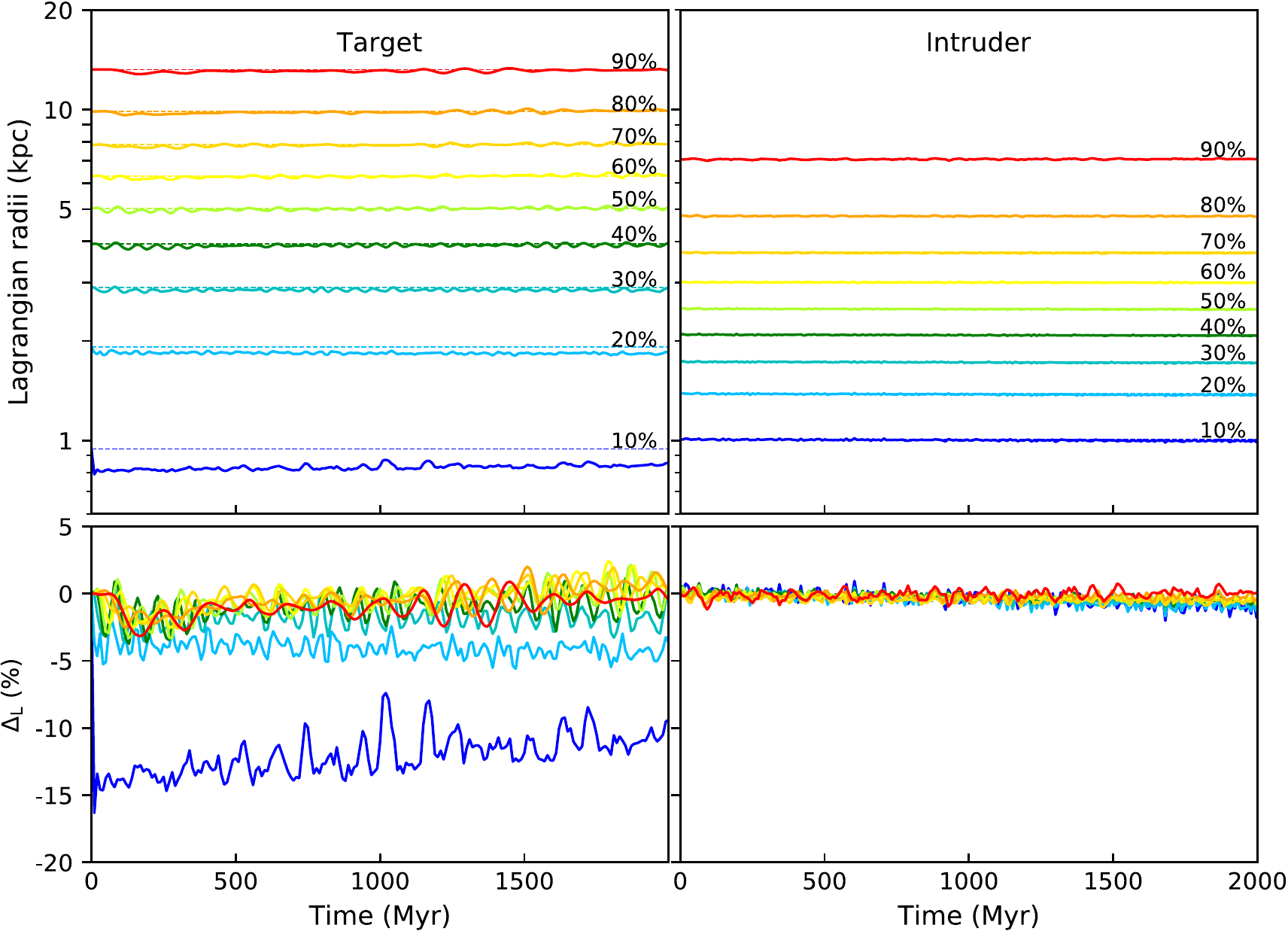}
\caption{Upper panels: the evolution of Lagrangian radii of the target galaxy (upper left-hand panel) and the intruder galaxy (upper right-hand panel). Lower panels: the change of Lagrangian radii compared to that at time $T=0$, $\Delta_L = [L_{\rm T=0}-L(T)]/L_{\rm T=0}$.}\label{lr}
\end{figure*}  

The projected mass density profiles of the disk (solid curves) and the bulge (dashed curves) of the target galaxy are shown in Fig. \ref{den2d}. After a free evolution of $2~\Gyr$, the projected density of the disk component increases within $0.3h \approx 1\kpc$, and the maximal increment is up to about a factor of $2$ within $0.1h=0.35\kpc$. In addition, there are tiny oscillations for the projected disk density in the outer region, where $R > h=3.5\kpc$. We also study the evolution of Lagrangian radii (i.e., the spherical radii where different fractions of masses are enclosed), increasing from $10\%$ to $90\%$ in steps of $10\%$, for the stellar component in the upper left-hand panel of Fig. \ref{lr}. These Lagrangian radii are denoted as $L_{\rm 0.1},~L_{\rm 0.2},...,L_{\rm 0.9}$. The radius $L_{\rm 0.1}$ relaxes from $\approx 0.9~\kpc$ to $\approx 0.8~\kpc$ at the very beginning of free evolution and oscillates around $ 0.8~\kpc$ in $2~\Gyr$ with an amplitude of roughly $\pm 0.05~\kpc$. Other Lagrangian radii do not show remarkable signs of evolution. The evolution of Lagrangian radii can more easily be found from the deviation of the Lagrangian radii of the ICs, which is defined as $\Delta_L = [L_{\rm T=0}-L(T)]/L_{\rm T=0}$. The $\Delta_L$ for each Lagrangian radii are shown in the lower panels of Fig. \ref{lr}, with the same colors used in the upper panels. For the target galaxy (the lower left-hand panel), the radius $L_{\rm 0.1}$ reduces by around $15\%$ at the very beginning and then slowly increases with time. At $T=2\Gyr$, the radius $L_{\rm 0.1}$ is about $10\%$ lower than that at $T=0$. The radius $L_{\rm 0.2}$ decreases by about $5\%$ within $2\Gyr$. The evolution of $L_{\rm 0.1}$ and $L_{\rm 0.2}$ implies that the core region of the target galaxy compresses, which agrees with the higher density in small radii of Fig. \ref{den2d}. Other Lagrangian radii oscillate around the original Lagrangian radii with an amplitude of $3\%$. The evolution of the central density profile indicates that the disk model is unstable. Since neither the bulge components nor the dark matter halos in the ICs are compact enough, the bulking instability appears in the free evolutions. Disk stars fall into the center of a galaxy and form an additional pseudo-bulge.
After $2~\Gyr$, the model is stabilized. Further investigations of the model instability are beyond the scope of this paper. Moreover, we also test the stability of the intruder galaxy and show the evolution of the Lagrangian radii in the right-hand panels of Fig. \ref{lr}. The intruder galaxy is stable. We will use the models freely evolved for a timescale of $2~\Gyr$ as the progenitors for the simulations of the colliding systems.

\subsection{Parameters for the Colliding Galaxy Pair}\label{sec-parameters}
To generate a collisional ring through off-center orbits, we will make a general analysis of the following parameters at first: (i) the initial relative velocity, $\vzero$, of the center of mass (CoM) of the target galaxy and the CoM of the intruder galaxy; (ii) the inclination angle, $\theta$; (iii) the impact parameter, $b$; and (iv) the total mass and scale length of intruder galaxies.


\begin{figure}
\centering  
\includegraphics[width=85mm]{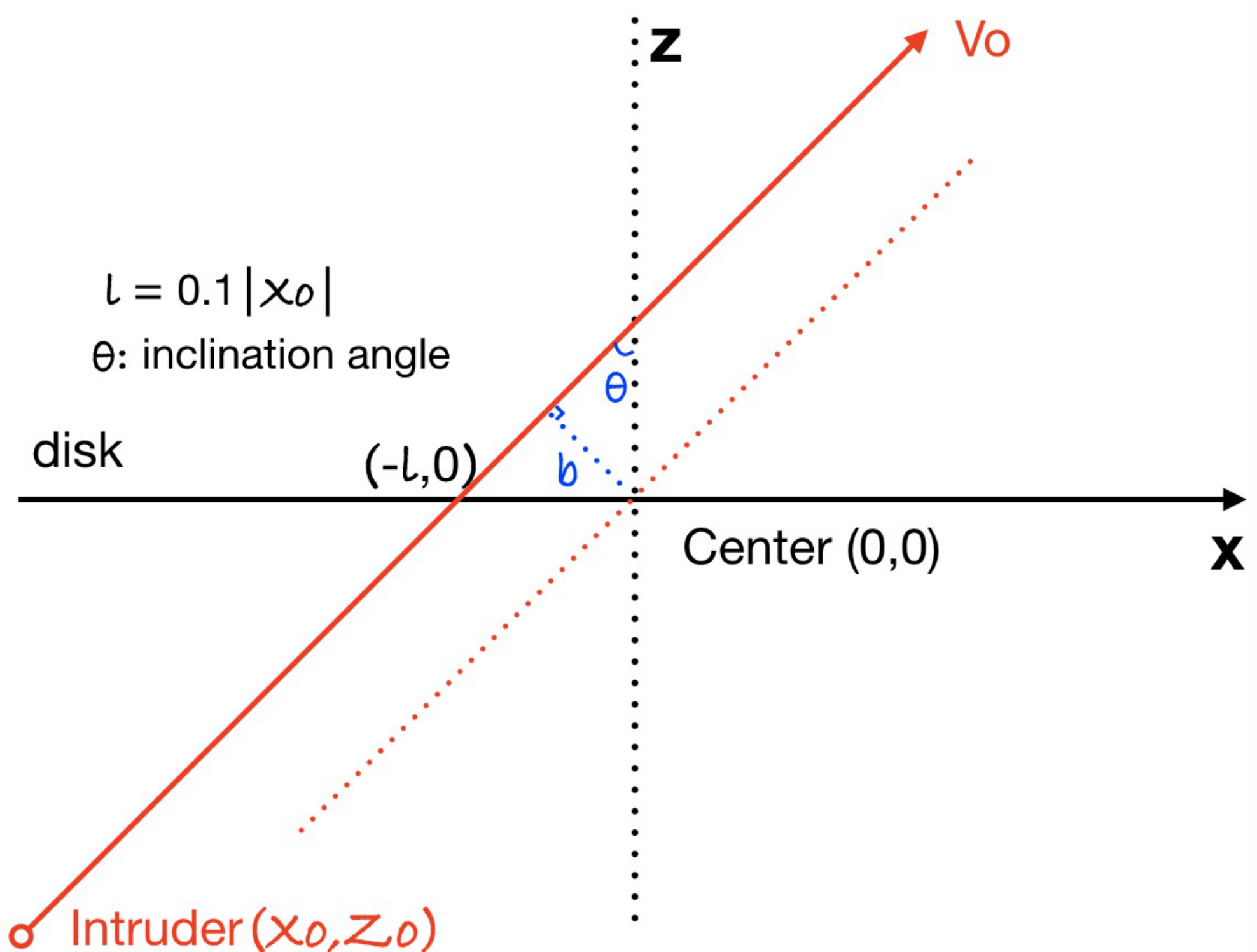}
\caption{The schematic diagram of the orbital parameters for the colliding galaxy pair.}
\label{schematic}
\end{figure}  

The CoM of the disk galaxy is located at the origin point of a Cartesian coordinate $(x,~y,~z)$ and the disk plane coincides with the $x$-$y$ plane. The initial position of the intruder galaxy is placed at $(x_0,~y_0,~z_0)=(-10.0h,~0.0,~z_0)$, where $h$ is the scale length of the disk model. We define a deviation scale length, $l$, to measure the distance of the original point and the intersection of $\vzero$ and the disk plane. The value of $l$ is fixed at $0.1x_0$ and the intersection is on the same side of the initial position of the intruder galaxy. A schematic diagram of the orbital parameters for the colliding galaxy pair is presented in Fig. \ref{schematic}. When the value of $\vzero$ is chosen, the initial position of the intruder galaxy (i.e., $z_0$), the impact parameter ($b$) and the inclination angle ($\theta$) are determined as follows,
\bey \label{z0b}
\theta &=& {\rm tan}^{-1} \left(\frac{v_{\rm x0}}{v_{\rm z0}}\right),\nonumber\\
z_0 &=& {\rm sign}(x_0)\cdot \frac{v_{\rm z0}}{v_{\rm x0}}(|x_0|-l)=0.9x_0\frac{v_{\rm z0}}{v_{\rm x0}},\\
b &=& \frac{l}{\cos \theta}.\nonumber
\eey

Since these above parameters have been broadly studied, we simulate the formation of CRGs to select the most suitable orbit to further study how $B/D$ affects the dynamics of the CRGs. The details of the model parameters and the reproduced results, including the projected stellar density contours of the CRGs, are presented in Appendix \ref{appendix-parameters}.

\subsubsection{Timescales of the CRGs}\label{Times_CRGs}

\begin{figure}  
\centering  
\includegraphics[width=85mm]{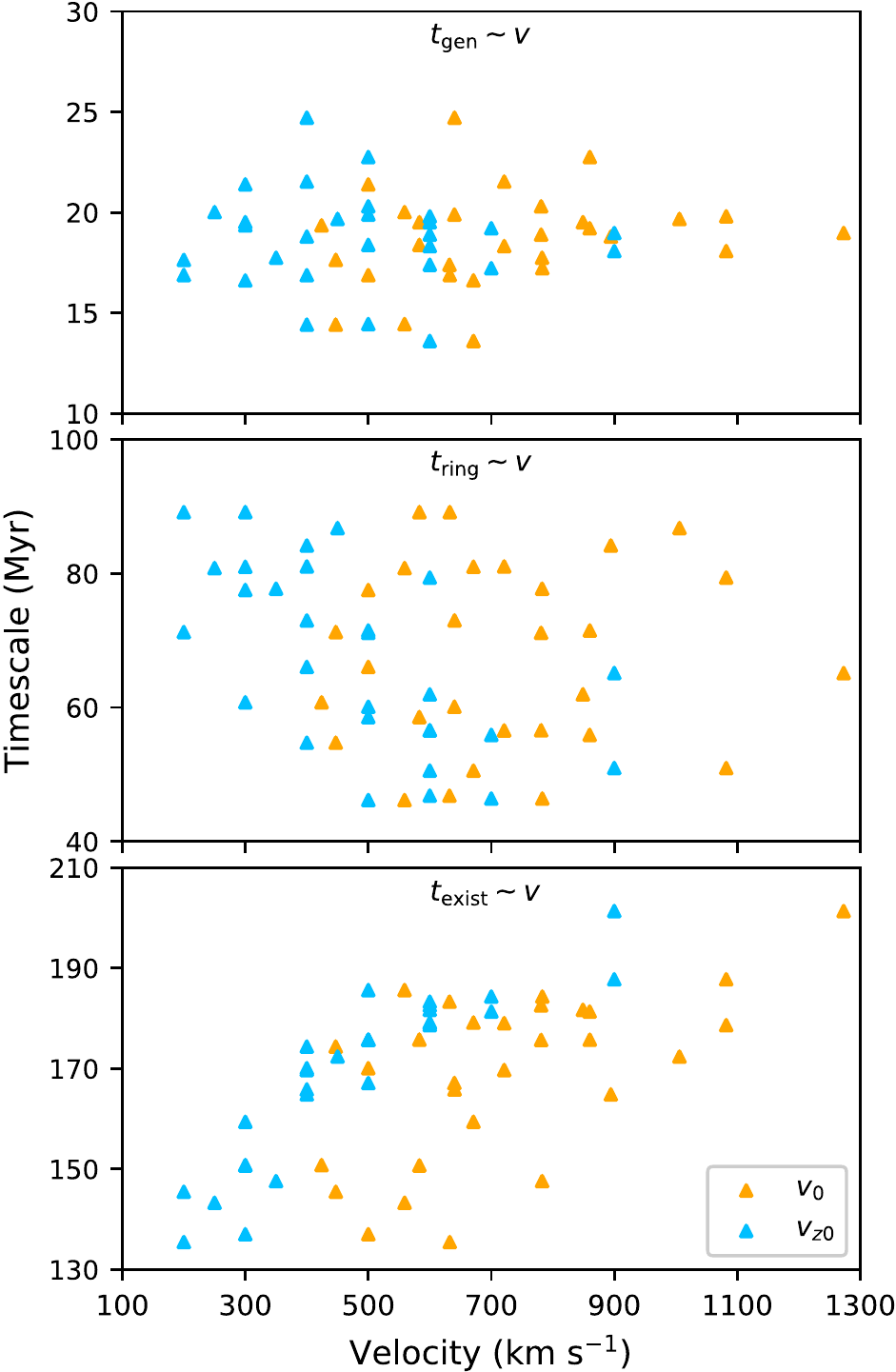}
\caption{The three panels represent the relations between the timescales ($t_{\rm gen}$ in the top panel, $t_{\rm ring}$ in the middle panel panel and $t_{\rm exist}$ in the bottom panel) and the initial relative velocities (total initial relative velocity $\vzero$ with orange regular triangle symbols and the impact velocity's component vertical to the target disk, $v_{\rm z0}$, with cyan triangle symbols). 
}
\label{v_Time}
\end{figure} 

The time scales $t_{\rm{gen}}$, $t_{\rm{ring}}$, $t_{\rm{exist}}$ denote the formation, the propagation out to $5h$, and to the edge of the disk galaxy. The time for two galaxies to encounter is recorded as $t_{\rm coll}$, which is obtained by linearly interpolating simulational time between two nearby snapshots. In this time interval, the sign of $z$ for the intruder galaxy changes for the first time. $t_{\rm{gen}}$ is defined as the time interval from $t_{\rm coll}$ to the time for a ring structure to propagate to $x\approx 2h=7\kpc$ (on the direction of positive $x$-axis, and the same definition for directions of the rings hereafter). We remind the reader that $h$ is the scale length of the disk for the initial target galaxy (see Table \ref{galmodels}). $t_{\rm ring}$ is the time interval from $t_{\rm coll}$ to the time for a ring to spread out to $x=5h$. For a target galaxy model, we choose a radius of $8.5h=30\kpc$ to specify the edge of the galaxy, which encloses over $99\%$ of the total stellar mass. The timescale for the density peak of a ring to spread to the edge of the disk galaxy, $x=8.5h$, is represented by $t_{\rm exist}$, which is comparable to the existing time of the ring. In summary, the relevant timescales are defined as follows,

\bey \label{eq-time}
t_{\rm{gen}}&=&t_{\rm{2h}}-t_{\rm{coll}},\nonumber \\
t_{\rm{ring}}&=&t_{\rm{5h}}-t_{\rm{coll}},\\
t_{\rm{exist}}&=&t_{\rm{8.5h}}-t_{\rm{coll}}.\nonumber
\eey

The values of these timescales are listed in Table \ref{orbitpara}, which can be found in Appendix \ref{appendix-parameters}. The relations between the timescales and the initial relative velocities are shown in Fig. \ref{v_Time}. A ring forms within $13-25\Myr$ after the collision, and propagates to $5h$ in a timescale less than $100~\Myr$. The timescale of $t_{\rm gen}$ does not correlate neither with $v_{\rm z0}$, nor with $v_0$. The propagation timescale, $t_{\rm ring}$, anti-correlates with $v_{\rm z0}$. However, there is no correlation between $t_{\rm ring}$ and $v_0$. This indicates that the propagation speed of the ring is related to $v_{\rm z0}$, the vertical component of initial impact velocity. Larger $v_{\rm z0}$ induces more rapid propagation of the ring, which arises from larger momentum of the higher speed intruder galaxy. However, this does not mean that the perturbation from the high speed intruder galaxy is stronger. Actually, the morphology of the CRGs in Appendix \ref{appendix-parameters} is less clear in the case of a higher speed encounter because the perturbation is an integrated effect of gravitation.

The rings exist for a timescale in the range of $[130,~210]~\Myr$. $t_{\rm exist}$ correlates with the initial relative velocity, $v_0$, and also with the $z$-component of the initial relative velocity, $v_{\rm z0}$. Higher $v_0$ and $v_{\rm z0}$ correspond to longer $t_{\rm exist}$. An immediate indication is that for a high speed encounter, the propagation slows down in a later stage of the ring because the time of gravitational interaction between the two galaxies is shorter and the accumulative perturbation from the intruder galaxy is weaker. The dispersion for the correlation between $t_{\rm exist}$ and $v_0$ is larger than that between $t_{\rm exist}$ and $v_{\rm z0}$ because the $v_{\rm z0}$ component of initial impact velocity is directly relevant to the formation of the ring structure.

Moreover, the mass of the intruder galaxy anti-correlates with the timescales of $t_{\rm gen}$, $t_{\rm ring}$ and $t_{\rm exist}$, while the scale length of the intruder galaxy correlates with the three timescales (see Table \ref{Intruder,time} in Appendix \ref{appendix-parameters}). Therefore, the formation, propagation and disappearance of the ring are more rapid for a more massive intruder galaxy and for a smaller sized the intruder galaxy. \citet{Struck-Marcell_Lotan1990} found that the velocity of a ring is proportional to the mass of the intruder galaxy, which agrees with our anti-correlation between the timescales of the ring and the mass of the intruder galaxy.

\subsubsection{Suitable Orbits for Testing $B/D$ as a Free Parameter}
The projected distances between SDSS J1634+2049 and its companion galaxies, C1 and C2, are $35.7\kpc\approx 10h$ and $44.0\kpc \approx 13h$, respectively. However, the luminosity of C2 galaxy is too low to get a spectrum \citep{liu2016sdss}. Here, we consider the case where the ring is a product of collision between SDSS J1634+2049 and C1. The ring structure is about $16\kpc \approx 5h$ to the galactic center of the host galaxy. We select the orbits from the models presented in Appendix \ref{appendix-parameters}. On these selected orbits, clear ring structures form in the target galaxy models at a radius of $5h$. At this time, we select models where the separation of the two galaxies on the disk plane of the target galaxy is around $(10\pm 2)h$. This position of ring and the separation of the two galaxies can mimic the observed properties of SDSS J1634+2019. There are three orbits satisfying the criteria; i.e., the orbits of models V17, V22 and V27 in Table \ref{orbitpara}. In other orbits, it is hard to reproduce both the separation of the two galaxies and the position of the ring at the same time step. Therefore, the three orbits will be used in the following simulations. 


\section{CRGs with Different $B/D$ of the Target Galaxy}\label{sec-bd}
\begin{figure*}
\centering
\includegraphics[width=135mm]{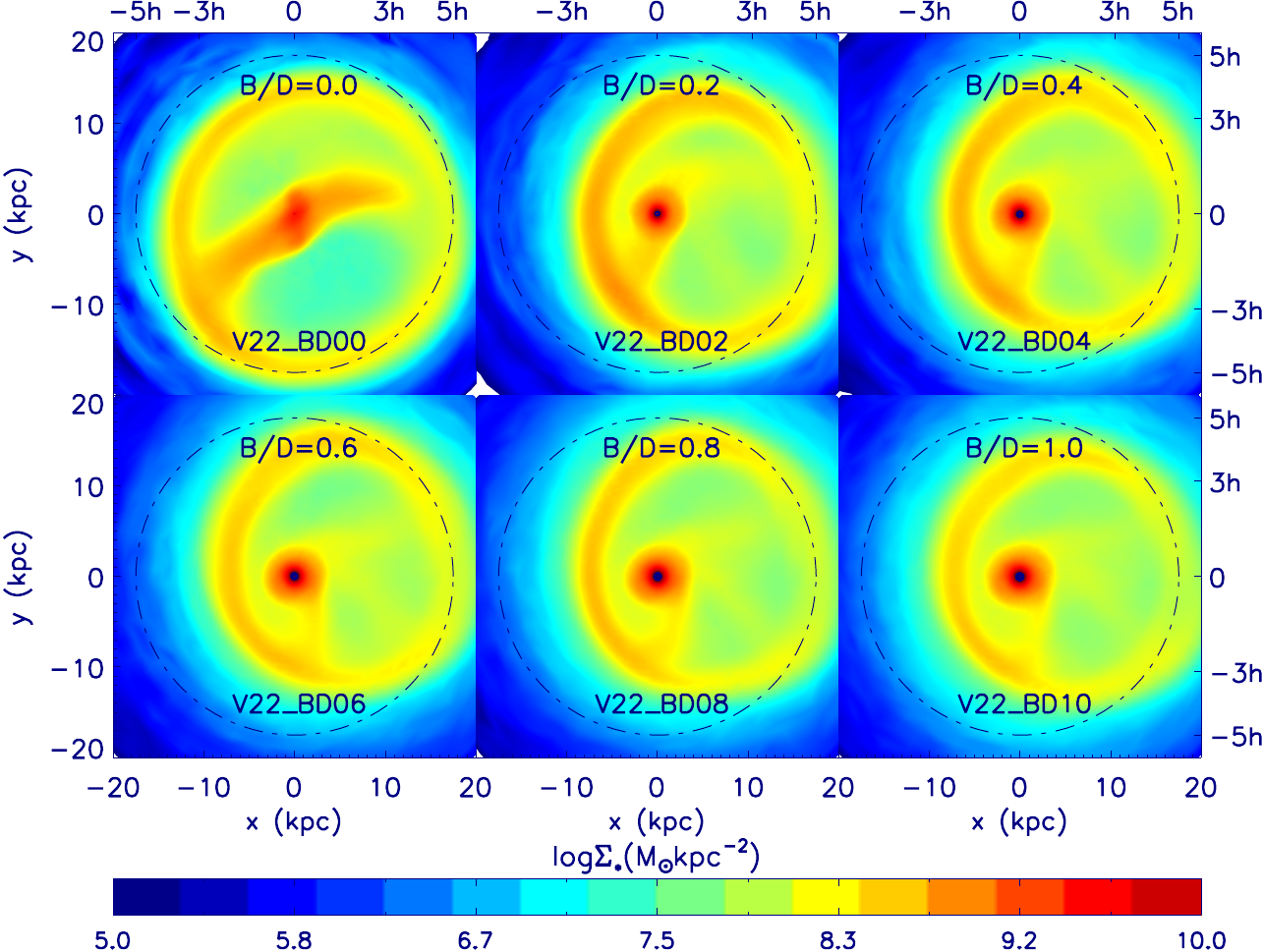}
\caption{The contours of the projected stellar-mass density of the ring galaxies models $V22\_BD00-V22\_BD10$ with $B/D$ rising from $0.0$ to $1.0$ with an interval of $0.2$, when the ring structure propagates out to $x \approx 5h$ (dashed circle).}
\label{contourBD47}
\end{figure*}

In observations, the $B/T \leq 0.2$ for $69\%$ bright spiral galaxies in a sample of $143$ bright and massive Sa-Sb Hubble type galaxies \citep{weinzirl2009bulge}. Moreover, $\approx 70\%$ of the edge-on disk galaxies are bulge-less or disk-dominated galaxies in a complete and homogeneous sample of $15127$ edge-on disk galaxies in the Sloan Digital Sky Survey (SDSS) Data Release 6 \citep{kautsch2009edge}. Meanwhile, disks form in CDM halos in the standard $\Lambda$ cold dark matter ($\Lambda$CDM) framework, while classical bulges form through mergers of disks and pseudo-bulges grow through intrinsic disk instabilities. Recently, baryonic feedback from supernovae, active galactic nuclei (AGNs) or accretion of cold gas from cosmic filaments have been incorporated into cosmological simulation and used to argue that realistic disk galaxies can form through a quiet merger \citep{agertz2010formation, guedes2011forming, aumer2013towards, marinacci2014diffuse}. However, in these models, the values of $B/T$ are generally larger than $0.3$ for massive spiral galaxies with a mass over $6\times 10^{10}\msun$. The disagreement of $B/T$ between observations and simulations is one of the main challenges in the $\Lambda$CDM cosmology. Since the formation of CRGs is a natural result of the interaction between galaxies, varying the structure of the parent galaxy might give rise to different morphology, kinematics and dynamics of the ring. Therefore, the ratio, $B/T$, serves as an important probe to test the hierarchical galaxy formation scenario. Unfortunately, little work has been done to investigate how the formation of the CRGs relies on the structure of the progenitor. Here, we will test the formation of CRGs by tuning the values of $B/D$, which is analogous to $B/T$, for the progenitors. 

The values of $B/D$ that are used in this paper range from $0.0$ to $1.0$, with an interval of $0.1$, and the models are denoted as from $V17\_BD00$ to $V17\_BD10$, from $V22\_BD00$ to $V22\_BD10$ and from $V27\_BD00$ to $V27\_BD10$, respectively, where the last two numbers in the names of models indicate the values of $B/D$. Other parameters are exactly the same as those of models V17, V22 and V27 in Table \ref{orbitpara}. The total simulation time for each set of model is $t_{\rm ring}$ in Table \ref{orbitpara}. We remind the reader that the stellar masses and the DM halo masses of the models are fixed at the values of $1.34\times 10^{11}\msun$ and $2.68\times 10^{12}\msun$. The particle numbers of the disk, bulge and DM halo components in a target galaxy are $4\times 10^{6}$, $2\times 10^{6}$ and $2\times 10^{5}$, which are the same as those listed in Table \ref{galmodels}. The particle masses of the bulge and disk are changed by the varying $B/D$. 

\subsection{Morphology}
For a given $B/D$, the morphologies of three sets of models are very similar, thus we only present the morphology of one set of CRGs; i.e., models on the orbit of V22, in Fig.\ref{contourBD47}. In this set of models (e.g., models on the orbit of V22), when the values of $B/D$ are $0.0-0.4$, corresponding to a bulge-less disk and a disk-dominated galaxy, the ring structures are the clearest in the set of models. 
Moreover, there is an apparent trend where the ring structures become less and less clear with increasing $B/D$. However, to confirm this, it is necessary to carry out more quantitative analysis. For instance, one can calculate the projected density profiles for the ring galaxies as discussed in Section \ref{B/D,Mass}.

  Moreover, since the stellar masses of the models are fixed, a smaller value of $B/D$ always leads to larger disk mass. More massive galactic disks form more distinct rings, while less massive disks form fainter rings. We will show that the ring structure is closely related to the value of $B/D$, and a smaller $B/D$ leads to a clearer ring structure. However, the ring structure is nearly independent of the disk mass. This can be verified by studying the image of the morphology based on the models with fixed disk mass, which is presented in Sec. \ref{fixdisk}.
  

\subsection{Mass Density Profiles of CRGs}\label{B/D,Mass}

\begin{figure*}  
\centering  
\includegraphics[width=140mm]{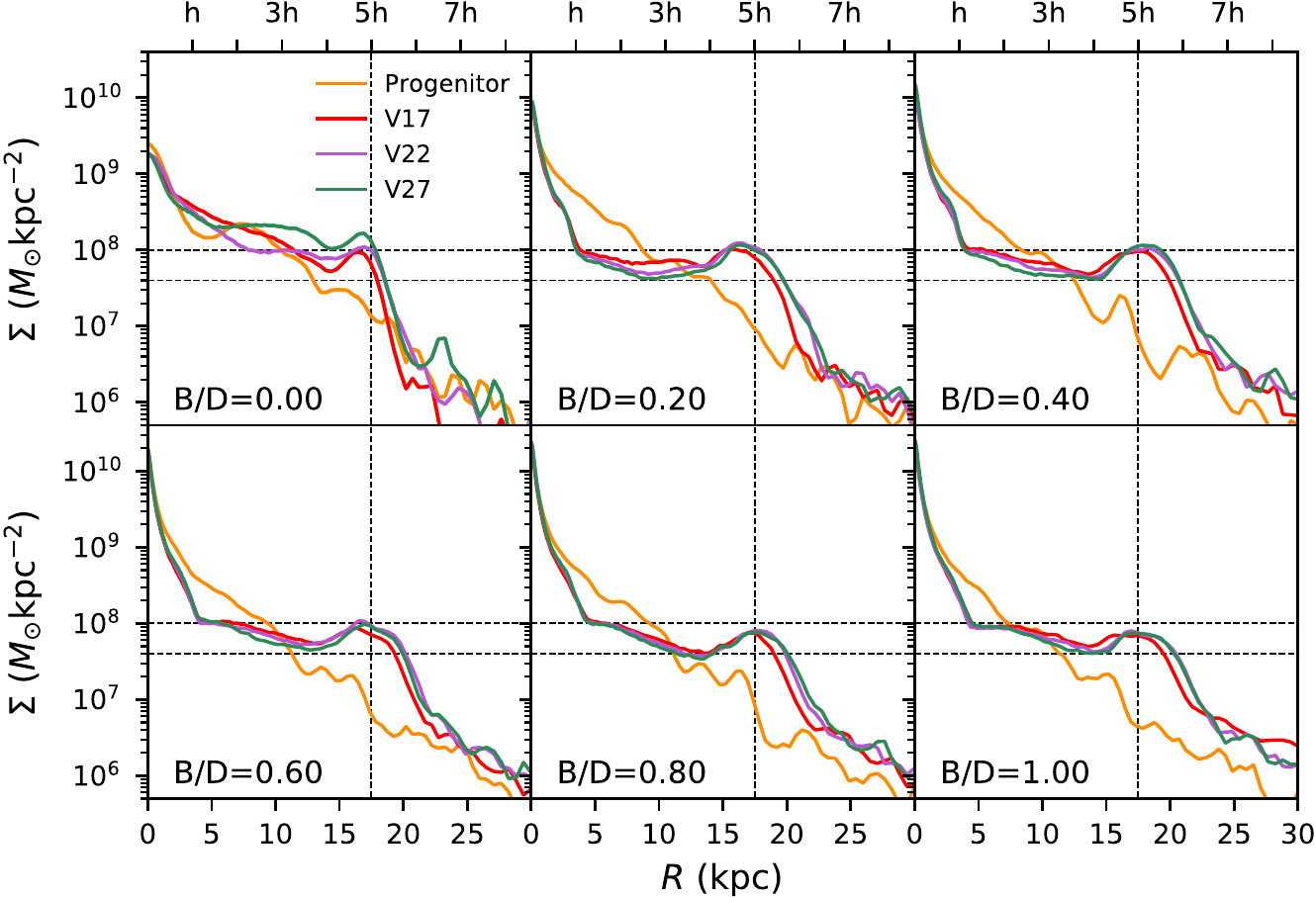}
\caption{Projected stellar-mass density profiles of the CRG with $B/D$ ranging in $[0.0,~1.0]$ with an increasing interval of $0.2$. The density profiles of the progenitors (orange) are shown in the figure and the density profiles of the three CRGs with different orbital parameters are presented with different colors: red (V17), purple (V22) and green (V27). The vertical dashed line in each panel marks the radius of $5h$, while the two horizontal dashed lines mark the projected density of $4\times 10^7\msun \kpc^{-2}$ (lower line) and $1\times 10^8\msun \kpc^{-2}$ (upper line). The ticks of x-axis are in units of $h$ (top) and of $\kpc$ (bottom).
}
\label{BD_mr}
\end{figure*}

The projected stellar density profiles on the x-axis of the ring galaxy models with the values of $B/D$ ranging in $[0.0,~1.0]$ are shown in Fig. \ref{BD_mr}. The projected density of the progenitors (orange curves) and of the CRGs at the time of $t_{\rm ring}$ on three different orbits (red curves for the orbit of V17, purple for the orbit of V22 and green for the orbit of V27) are presented in the figure. First, we find that the disk is not stable for the bulge-less ($B/D=0.0$) model. After a free evolution for $2\Gyr$, a density peak appears at the radius of $x\approx 2h$ for the progenitor. Therefore, a new structure is formed at the radius of $2h$, which is related to the disk instability. In a pure-disk galaxy, a bar structure develops from the secular evolution of the thin disk if the dark matter halo is not very concentrated in the center of the pure-disk galaxy \citep[e.g.][]{shen2010our}. For non-zero values of $B/D$, no such density peaks are generated after the stability test for the ICs. Hence, the bulge component stabilizes a disk galaxy model, which is similar to the effect of a concentrated dark matter halo center \citep{sotnikova2005bending}.

For a model with $B/D=0.0$, after the collisions, the density peaks at $2h$ become milder for the CRGs and shift to larger radii compared to that of the progenitors. A possible reason for this is that the matter in the pseudo-bulge structure is heated through the collision and has a more extended structure. The surface density contour in the upper left-hand panel of Fig. \ref{contourBD47} confirms an extended bar structure in the CRG. Moreover, there is another density peak propagating out to $x\approx 5h$, corresponding to the ring structures in the upper left-hand panel of Fig. \ref{contourBD47}. The density peaks reveal the positions of the rings on the direction of x-axis.

For a target galaxy model containing a bulge and a disk ($B/D>0.0$), a clear density peak emerges and then propagates outwards the disk plane after the collision. When the density peak moves to $R=5h$, the projected stellar density, $\Sigma$, at the radius of $5h$ is approximately a factor of $2$-$3$ larger than that of the background disk plane in the inner side. $\Sigma(R)$ substantially decreases in the outer side of the peak. Comparing to the progenitors, $\Sigma (5h)$ is about an order of magnitude larger after collision. Moreover, $\Sigma(R)$ decreases with growing $R$ within $R\in [0.3h,~3h]$. The stellar particles within $R\in [0.3h,~3h]$ of the progenitor contribute to the formation of the ring. 

Moreover, the values of $\Sigma(R)$ at $R=5h$ are the largest for models with $B/D \in [0.0,~0.6]$. In these models, $\Sigma (5h)$ exceeds or reaches up to $10^8 \msun \kpc^{-2}$. As $B/D$ is above $0.6$, $\Sigma(5h)$ is reduced for all the models. In addition, for models with $B/D = 0.2$ and $0.4$, the density gaps near $4h$ are close to $4\times 10^7 \msun \kpc^{-2}$. This indicates that the projected density of the rings in these two models are about a factor of $2.5$ higher than that of the gaps inside the rings. For the models with $B/D=0.0$ or $B/D>0.6$, the ratios of the projected densities between the rings and the gaps are smaller. Thus, we see that the rings are the clearest for the models with $B/D=0.2$ and $0.4$.
This reflects the fact that, for a given total mass of the stars in a galaxy, a larger value of $B/D$ implies a smaller mass of the disk component. The ring structure generated from the disk is weaker in amplitude for a lighter disk. 
\subsubsection{Mass of the Rings}\label{rmass}
\begin{table}
\centering
\caption{Mass of Rings with Different Values of $B/D$ on Orbits of V17, V22 and V27} \vskip 0.2cm
\begin{threeparttable}
\begin{tabular}{lccccccc}
\hline
\hline
Orbit & V17 & V22 & V27 \\
\hline
$B/D$  &$M_{\rm{ring}}$  &$M_{\rm{ring}}$  &$M_{\rm{ring}}$ \\
$\space$ & ($10^{10}\Msun$) & ($10^{10}\Msun$)  & ($10^{10}\Msun$) \\
\hline
0.00 & 4.45 & 3.47 & 3.68 \\
0.10 & 3.26 & 3.00 & 3.67 \\
0.20 & 3.22 & 2.96 & 3.49 \\
0.25 & 3.13 & 2.93 & 3.31 \\
0.30 & 2.98 & 2.75 & 3.26 \\
0.40 & 3.03 & 2.87 & 3.15 \\
0.50 & 2.99 & 2.79 & 2.99 \\
0.60 & 2.88 & 2.77 & 3.02 \\
0.70 & 2.64 & 2.49 & 2.74 \\
0.80 & 2.48 & 2.34 & 2.64 \\
0.90 & 2.39 & 2.26 & 2.56 \\
1.00 & 2.35 & 2.19 & 2.45 \\
\hline
\end{tabular}
\label{BD,Ringmass}
\begin{tablenotes}
\item 
\begin{flushleft}
\textbf{Note.} The models with $B/D=0.25$ are CRGs formed from the controlled progenitor models that were used for the stability tests in Sec. \ref{ics}.
\end{flushleft}
\end{tablenotes}
\end{threeparttable}
\end{table}
The ring particles are defined as the particles in the region where $r\in[4h, 6.5h]$ and with positive radial velocities (i.e., $v_{\rm R}$, defined in Sec. \ref{seckin}). 
We list the masses of the rings that propagate out to $5h$ in Table \ref{BD,Ringmass}, with different $B/D$ on orbits of V17, V22 and V27. The masses of the rings are between $2.2 \times 10^{10}\msun$ and $4.5\times 10^{10}\msun$, which are approximately $16\%$-$33\%$ of the overall stellar masses of the progenitors. In general, rings at $R=5h$ are more massive for models with smaller $B/D$. This agrees with the fact that the amplitudes of the projected density peaks decrease at $5h$ with growing $B/D$ (Fig. \ref{BD_mr}). In addition, an orbit with lower relative velocity (i.e., orbit V17) forms a more massive ring because the time for gravitational interaction between the galaxy pair is longer and the cumulated perturbation is stronger. 

\subsubsection{S\'{e}rsic Index of the Central Nucleus}\label{sec-index_n}
\begin{figure}
\includegraphics[width=85mm]{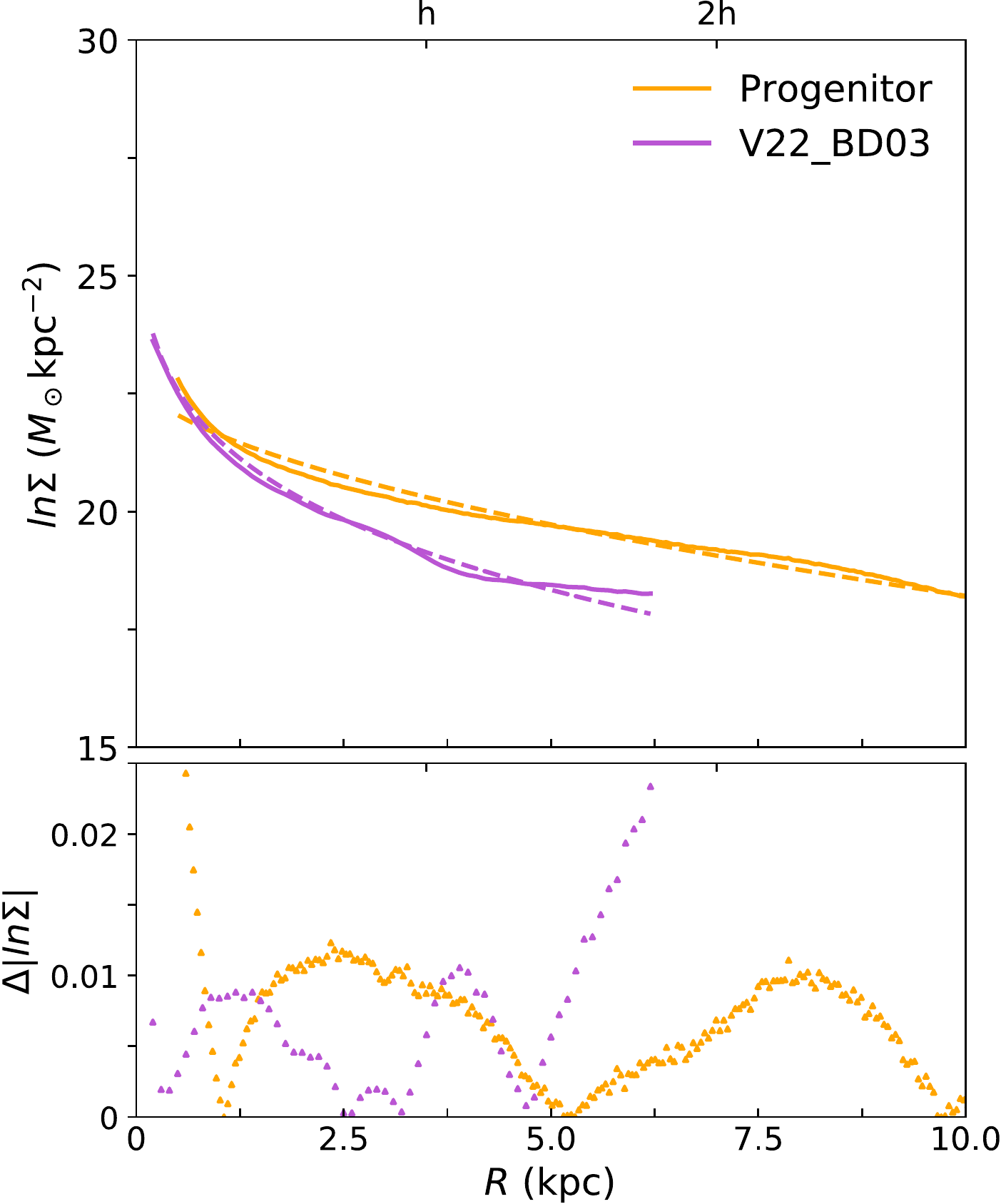}
\caption{Upper panel: S\'{e}rsic model fitting curves (dashed curves) for the progenitor galaxy with $B/D=0.3$ (orange curves, labelled as `progenitor') and for the galaxy after collision (purple curves labelled as `V22\_BD03'), and the surface density computed from the \textit{N}-body model is represented by solid curves. Lower panel: the residuals between the surface density from the \textit{N}-body model and the S\'{e}rsic model fitting. The units for the horizontal axes of the two panels are in both $\kpc$ (bottom) and in $h$ (top).}
\label{NFit_ex}
\end{figure}

S\'{e}rsic's profile \citep{sersic1968formation, capaccioli1989photometry} has been extensively used to describe the light-profiles of the early-type galaxies (ETGs) and the bulges of spiral galaxies. The S\'{e}rsic's surface density profile is given by the following expression \citep{graham2003new} 
\bey
\Sigma(R)&=& \Sigma_e \exp \left\{ -b_n \left[ \left(R\over R_e \right)^{1/n}-1 \right] \right\}, \nonumber \\
b_n&\approx& 1.9992n-0.3271, \space\space\space{\rm for} 1\lesssim n\lesssim 10,
\label {eq3-1}
\eey
where $n$ is S\'{e}rsic index and $\Sigma_e$ is central surface density at half-light radius, $R_e$. The S\'{e}rsic index, $n$, describes the shapes of density profiles for galaxies. For an exponential disk profile, $n=1$, while for the de Vaucouleurs model, $n=4$.

We now use the S\'{e}rsic's profile to study the shapes of the projected stellar density profiles of the model galaxies. The truncation radius is $30\kpc\approx 8.5h$ for the ICs of the target galaxies and for the progenitors. For the galaxy models at the time of $t_{\rm ring}$, the fitting range of the galaxies are truncated at a radius of $3R_{\rm e}$ to exclude the projected density peaks (i.e., the rings). Here, $R_{\rm e}$ is the half projected stellar mass for the central remnant of a CRG and the central remnant is truncated at a radius of $10\kpc \approx 3h$, where a density gap appears between the disk and the ring. The stellar particles are binned radially with equal radial intervals of $0.1\kpc$. The radius of the innermost bin of data is much larger than the softening radius. In addition, we only calculate the region of the positive $x$-axis for a ring galaxy because the ring is lopsided and the density gap appears at a radius smaller than $10\kpc \approx 3h$ in the region of the negative $x$-axis. Practically, we use a logarithmic form of S\'{e}rsic's profile to represent the projected stellar density of the model galaxies,
\begin{equation}\label {logsersic}
\ln[\Sigma(R)]= \ln(\Sigma_{e})  -b_n \left[ \left(R\over R_e \right)^{1/n}-1 \right] .
\end{equation}
The surface stellar densities of a galaxy model (model V22\_BD03 in Table \ref{Nfit_table}) before (green solid curves) and after (purple solid curves) the collision are shown in the upper panel of Fig. \ref{NFit_ex}, together with the S\'{e}rsic fitting profiles (dashed curves).  
 The residuals, $\Delta \ln [\Sigma (R)] \equiv \frac{|\ln [\Sigma_{\rm fit} (R)]-\ln [\Sigma (R)]|}{\ln [\Sigma (R)]}$, are shown in the lower panel of Fig. \ref{NFit_ex}, where $\Sigma_{\rm fit} (R)$ are the fitting values of surface density using Eq. \ref{logsersic}. The residuals are smaller than $2.5\%$ for this model. The projected stellar density of the model is well-fitted by the S\'{e}rsic profile. The parameters of S\'{e}rsic profile fitting for all the models with different values of $B/D$ are listed in Table \ref{Nfit_table}. The maximal residuals for all the models are less than $6\%$, hence the central regions of the CRGs can be fitted-well by using the S\'{e}rsic profile.

\begin{table}
\centering
\caption{Parameters of S\'{e}rsic Model Fitting for the ICs, the Progenitors and the Central Nuclei of CRGs}  \vskip 0.2cm
\begin{threeparttable}
\begin{tabular}{lcccccccccc}
\hline
\hline
$B/D$ & Models & $n$ & $R_{e}$ & $\Sigma_{e}$ & Max($\Delta \ln\Sigma$) \\
$\space$ & $\space$ & $\space$ &(kpc) & ($10^8\Msun\kpc^{-2}$) & (Percent) \\
\hline
0.00 & ICs        & 0.99 & 5.84 & 3.38 & 4.32 \\
0.10 & ICs        & 1.23 & 5.53 & 3.32 & 1.95 \\
0.20 & ICs        & 1.46 & 5.18 & 3.58 & 4.42 \\
0.25 & ICs        & 1.60 & 5.12 & 3.46 & 2.56 \\
0.30 & ICs        & 1.73 & 5.04 & 3.50 & 4.67 \\
0.40 & ICs        & 2.09 & 4.93 & 3.44 & 4.38 \\
0.50 & ICs        & 2.33 & 4.72 & 3.53 & 4.98 \\
0.60 & ICs        & 2.68 & 4.66 & 3.47 & 4.74 \\
0.70 & ICs        & 3.15 & 4.66 & 3.34 & 3.56 \\
0.80 & ICs        & 3.38 & 4.44 & 3.56 & 3.71 \\
0.90 & ICs        & 3.81 & 4.41 & 3.47 & 3.64 \\
1.00 & ICs        & 4.18 & 4.38 & 3.40 & 3.80 \\
0.00 & Progenitor & 1.29 & 5.72 & 3.21 & 1.24 \\
0.10 & Progenitor & 1.44 & 5.55 & 3.17 & 2.69 \\
0.20 & Progenitor & 1.50 & 4.45 & 2.33 & 5.30 \\
0.25 & Progenitor & 1.47 & 5.17 & 3.38 & 4.93 \\
0.30 & Progenitor & 1.80 & 5.15 & 3.50 & 4.25 \\
0.40 & Progenitor & 2.60 & 5.26 & 2.92 & 5.24 \\
0.50 & Progenitor & 2.73 & 5.12 & 3.05 & 4.47 \\
0.60 & Progenitor & 3.12 & 5.23 & 2.85 & 4.80 \\
0.70 & Progenitor & 3.02 & 5.26 & 2.72 & 3.75 \\
0.80 & Progenitor & 3.72 & 4.20 & 4.16 & 3.90 \\
0.90 & Progenitor & 3.25 & 4.08 & 4.12 & 1.85 \\
1.00 & Progenitor & 3.64 & 3.95 & 4.10 & 3.01 \\
0.00 & V17\_BD00  & 2.20 & 6.57 & 1.23 & 1.86 \\
0.10 & V17\_BD01  & 3.53 & 3.12 & 1.65 & 4.58 \\
0.20 & V17\_BD02  & 3.73 & 3.13 & 2.03 & 4.26 \\
0.25 & V17        & 4.35 & 2.93 & 3.40 & 3.56 \\
0.30 & V17\_BD03  & 4.71 & 3.12 & 2.12 & 4.05 \\
0.40 & V17\_BD04  & 4.97 & 3.40 & 1.75 & 3.22 \\
0.50 & V17\_BD05  & 5.66 & 2.20 & 4.98 & 2.35 \\
0.60 & V17\_BD06  & 6.81 & 1.95 & 6.50 & 1.78 \\
0.70 & V17\_BD07  & 6.07 & 1.94 & 6.69 & 1.98 \\
0.80 & V17\_BD08  & 9.16 & 1.77 & 9.01 & 2.20 \\
0.90 & V17\_BD09  & 7.36 & 2.06 & 5.97 & 1.96 \\
1.00 & V17\_BD10  & 9.00 & 1.72 & 9.04 & 1.44 \\
0.00 & V22\_BD00  & 2.75 & 7.04 & 0.98 & 1.97 \\
0.10 & V22\_BD01  & 2.55 & 6.83 & 0.83 & 4.30 \\
0.20 & V22\_BD02  & 4.99 & 4.21 & 1.15 & 3.39 \\
0.25 & V22        & 4.95 & 2.71 & 3.91 & 2.70 \\
0.30 & V22\_BD03  & 4.86 & 2.10 & 5.79 & 2.34 \\
0.40 & V22\_BD04  & 6.19 & 1.89 & 7.09 & 1.89 \\
0.50 & V22\_BD05  & 6.97 & 1.85 & 7.26 & 2.48 \\
0.60 & V22\_BD06  & 6.10 & 2.10 & 5.93 & 2.47 \\
0.70 & V22\_BD07  & 6.29 & 1.92 & 7.10 & 1.59 \\
0.80 & V22\_BD08  & 7.97 & 1.85 & 8.35 & 1.80 \\
0.90 & V22\_BD09  & 7.83 & 2.04 & 6.31 & 2.03 \\
1.00 & V22\_BD10  & 7.55 & 1.91 & 7.43 & 1.45 \\
0.00 & V27\_BD00  & 2.08 & 7.05 & 0.93 & 3.35 \\
0.10 & V27\_BD01  & 3.77 & 3.13 & 2.00 & 4.66 \\
0.20 & V27\_BD02  & 4.52 & 2.52 & 4.35 & 2.69 \\
0.25 & V27        & 4.10 & 2.58 & 4.19 & 3.23 \\
0.30 & V27\_BD03  & 4.58 & 3.30 & 2.30 & 4.43 \\
0.40 & V27\_BD04  & 5.69 & 2.09 & 6.35 & 1.74 \\
0.50 & V27\_BD05  & 6.18 & 1.97 & 7.18 & 2.11 \\
0.60 & V27\_BD06  & 7.33 & 1.74 & 9.26 & 1.83 \\
0.70 & V27\_BD07  & 8.55 & 2.02 & 6.79 & 1.81 \\
0.80 & V27\_BD08  & 7.44 & 1.99 & 7.84 & 2.10 \\
0.90 & V27\_BD09  & 7.17 & 2.56 & 4.78 & 2.37 \\
1.00 & V27\_BD10  & 8.96 & 1.81 & 8.83 & 1.59 \\
\hline
\end{tabular}
\label{Nfit_table}
\begin{tablenotes}
\item 
\begin{flushleft}
\textbf{Note.} Column 1 lists the values of $B/D$ and column 2 shows the names of the models. Columns  3--6 present the S\'{e}rsic Index, $n$, the projected half mass radius, $R_e$, the central surface density within $R_e$, $\Sigma_e$, and the maximal residual of the S\'{e}rsic model fitting, max($\Delta\ln \Sigma$), respectively.
\end{flushleft}
\end{tablenotes}
\end{threeparttable}
\end{table}

\begin{figure}
\centering
\includegraphics[width=85mm]{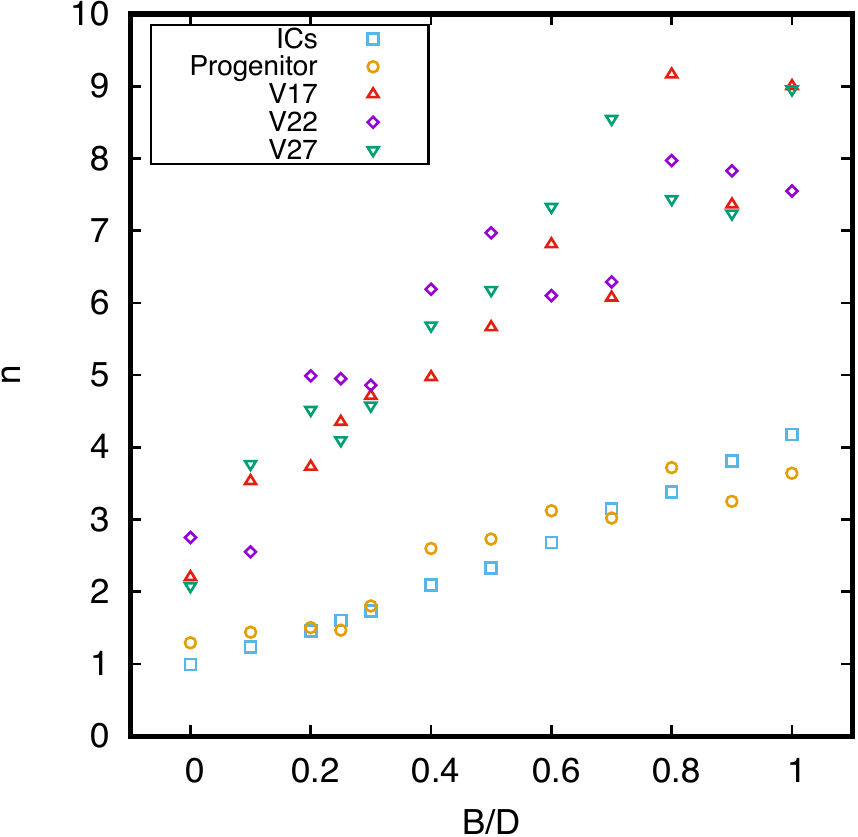}
\caption{S\'{e}rsic indices $n$ correlate with the values of $B/D$ in the model galaxies, including the ICs (cyan squares), the progenitors (orange circles) and the central remnants after collisions (red triangles, purple diamonds and green inverse triangles representing CRG models on the V17, V22 and V27 orbits, respectively). 
}
\label{BD_nfit}
\end{figure}

It has been found that the S\'{e}rsic indices of the final products increase in bulge-less galaxy-pair mergers (e.g., \citealt{gonzalez2005elliptical,  aceves2006sersic, naab2006surface}) or accretion of disk galaxy from satellites \citep{aguerri2001growth, eliche2005satellite, eliche2006growth}. In the process of CRG formation, the following question naturally arises: will the S\'{e}rsic index change significantly after the collision of the galaxy pair? In this section, we will study the relation between the S\'{e}rsic index and the initial values of $B/D$.

Fig. \ref{BD_nfit} shows the relation of the two parameters. For the ICs of the target galaxy, $n$ starts from 1.0 for the bulge-less disk galaxy model, increases with the growth of $B/D$ and ends up to $\approx 4.0$ when the masses of the bulge and the disk are equal $(B/D=1.0)$. After carrying out the stability test, the values of $n$ do not evolve significantly for the progenitor models. For a CRG at the time of $t_{\rm ring}$, the value of $n$ increases significantly compared to that of the progenitor. The value of $n$ increases the least for the CRG formed from the bulge-less target galaxy (model with $B/D=0.0$). On the orbits of V22 and V27, the values of $n$ increase by about $80\%$ at $t_{\rm ring}$. As a comparison, on the orbit V17, the value of $n$ increases by over a factor of $2$. When the target models contain a bulge component ($B/D > 0.0$), $n$ increases by a factor of $2$-$3$ after collisions. There is an almost linear relation between $n$ of the CRGs and the initial $B/D$. Apparently, the S\'{e}rsic index increases significantly after a minor collision event. The values of $B/D$ for the CRGs are larger than that of the final products in disk galaxy mergers \citep{aguerri2001growth,naab2006surface}. The non-equilibrium state of the CRGs might account for the larger values of $n$. 

Moreover, we find that $n\approx 4$ when the initial $B/D\in [0.1,~0.3]$. When the $B/D \geq 0.4$, $n>5.0$ for the CRGs at $t_{\rm ring}$. Therefore, for the progenitor galaxy with a minor bulge, the central remnant of the ring galaxy appears to be of early-type after an off-center minor collision. If a CRG is observed with a late-type central galaxy, then the progenitor must be bulge-less ($B/D<0.1$).

\subsection{Kinematics of the CRGs}\label{seckin}
To study the rotation and the radial motions of the stars in the CRGs, we calculated the average radial velocity, $v_{\rm R}(R)$ and the average azimuthal velocity, $v_{\phi}(R)$, of the disk semi-plane with positive $x$-axis. We also calculated the vertical velocity, $v_{\rm z}(R)$, for the particles with positive $x$-axis.
\bey\label{eq3-2}
v_{\rm R}=\frac{\sum_i m_i v_{\rm R,i}}{\sum_i m_i},\nonumber \\
v_{\rm \phi}=\frac{\sum_i m_i v_{\rm \phi,i}}{\sum_i m_i},\\
v_{\rm z}=\frac{\sum_i m_i v_{\rm z,i}}{\sum_i m_i},\nonumber
\eey
where $m_i$ is the mass of the $i^{th}$ particle in a grid cell on the disk plane (i.e., the $x$-$y$ plane) with positive $x$-axis. $v_{\rm R,i}$, $v_{\rm \phi,i}$ and $v_{\rm z,i}$ are the radial, azimuthal and vertical velocities of the $i^{th}$ particle in a cell.
\begin{figure*}
\centering
\includegraphics[width=170mm]{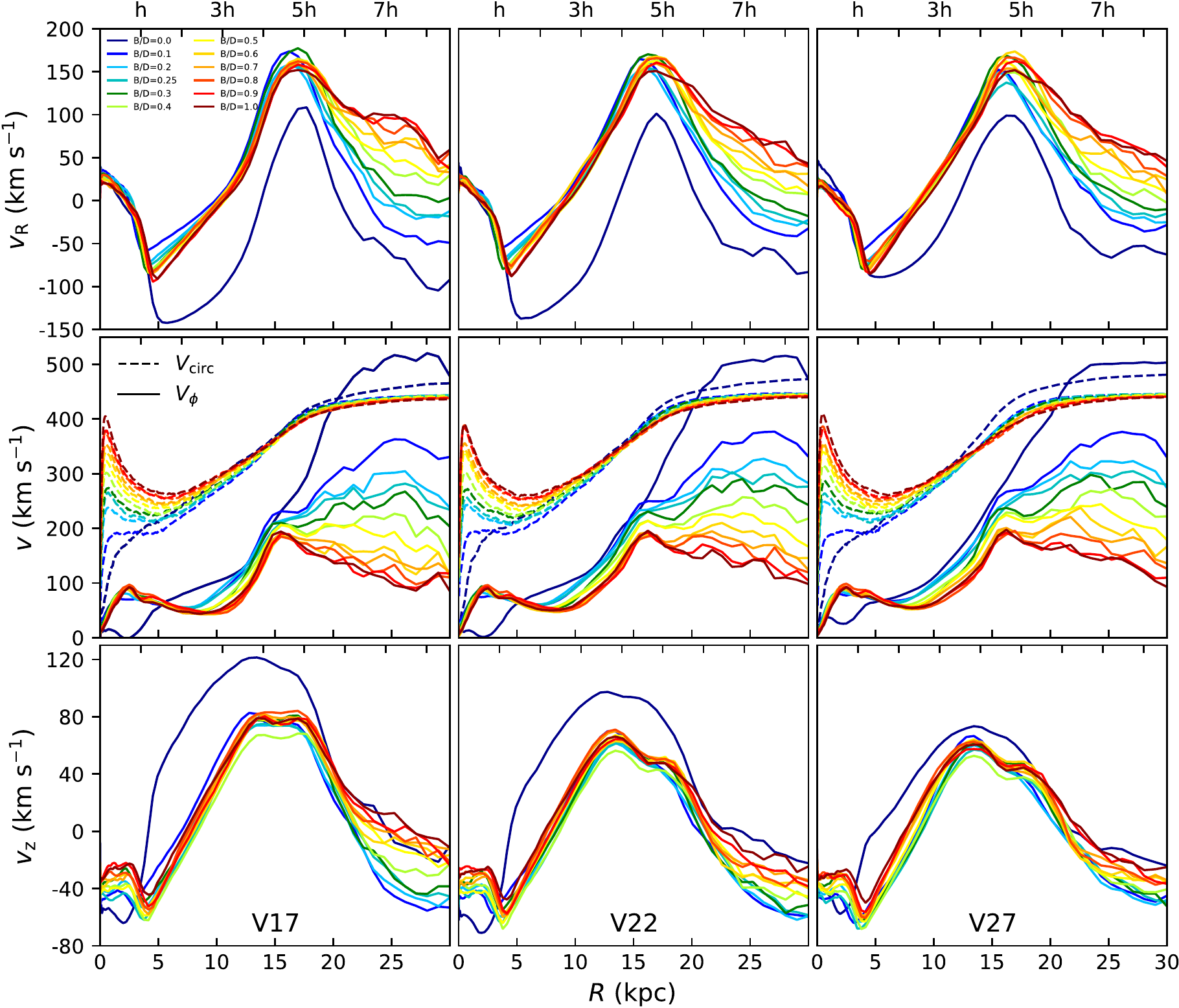}
\caption{The average radial velocities, $v_{\rm R}(R)$ (upper panels), the average azimuthal velocities, $v_{\rm \phi}(R)$, and the circular velocities, $v_{\rm circ}$, (solid and dashed curves in middle panels, respectively) of the CRGs formed from target galaxy models with different $B/D$ on different orbits (V17, V22 and V27 in the left-hand, middle and right-hand panels, respectively). The different values of $B/D$ for the models are distinguished by using different colors. The lower panels show the vertical velocity, $v_{\rm z}(R)$ of the CRGs.}
\label{BD_vr_vphi}
\end{figure*}

\subsubsection{Radial Velocities}
\begin{figure}
\centering
\includegraphics[width=85mm]{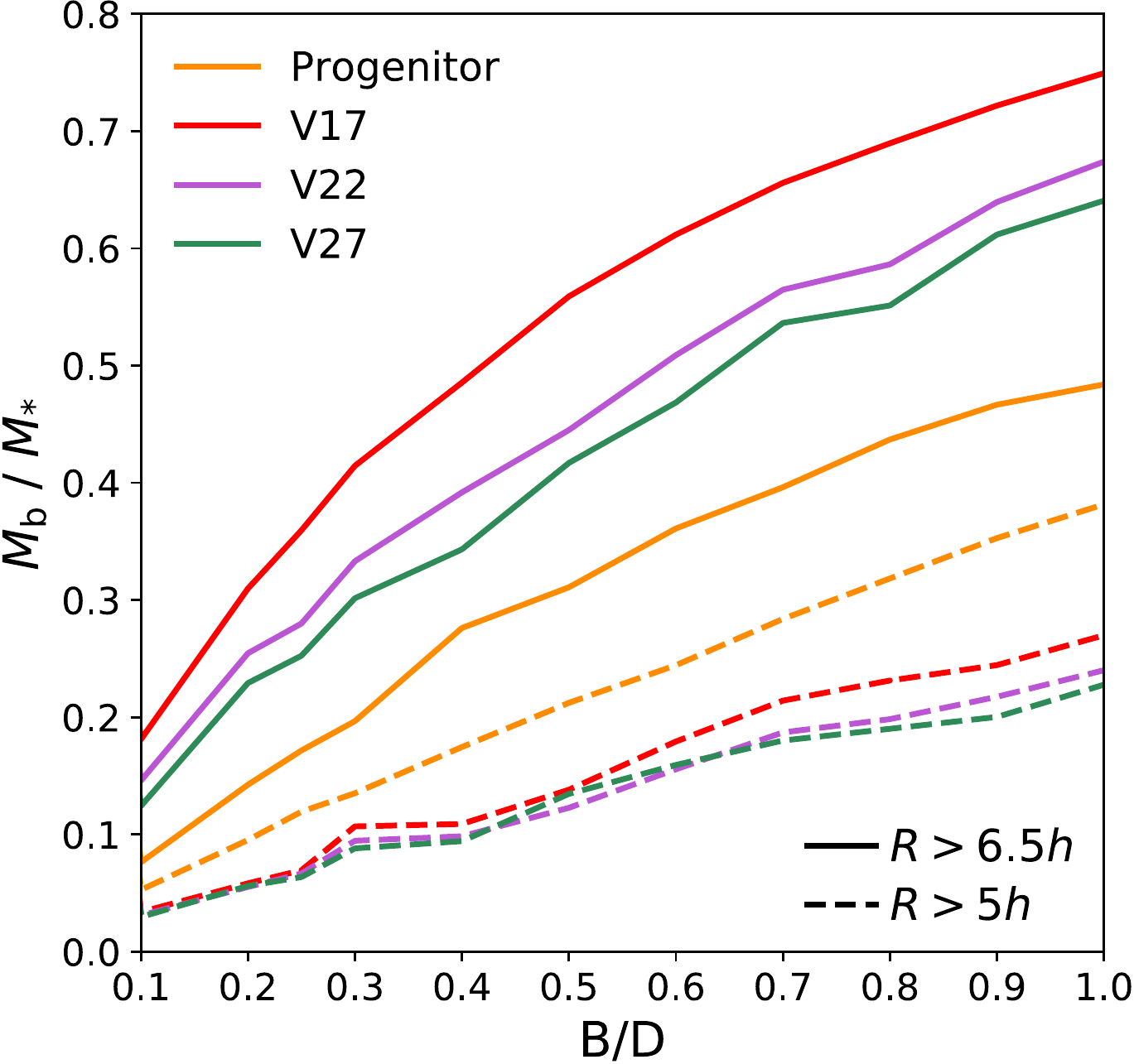}
\caption{The outer-bulge-to-outer-stellar-mass fractions in large radii (dashed curves for $R>5h$ and solid curves for $R>6.5h$) of the CRGs and of the progenitors with different $B/D$.
}
\label{bulgefrac}
\end{figure}	

The upper panels of Fig. \ref{BD_vr_vphi} show the $v_{\rm R}(R)$ profiles for CRG models formed on three orbits, namely orbit V17 (upper left-hand panel), orbit V22 (upper middle panel) and orbit V27 (upper right-hand panel). When a ring forms at a radius of $x=5h$, there is a positive peak for $v_{\rm R}(R)$ with an amplitude of around $100$-$170~\kms$ at the same radius of the density peak. The value of $v_{\rm R}(R)$ at $x=5h$ is the propagation speed of the ring, and $v_{\rm R}(R)$ is positive in the radii larger than that of the outer edge of the ring for the CRGs with $B/D>0.3$. Interestingly, we find that the amplitude of $v_{\rm R}$ outside of the outer edge of the ring ($R> 6.5h$) is an increasing function of the initial $B/D$. This implies that for a more bulge-dominated system undergoing a collision, the expansion velocity along the radial direction is higher outside of the outer edge of the ring. For the disk galaxies with a bulge, different values of $B/D$ lead to distinct $v_{\rm R}(R)$ profiles in large radii (i.e., outside of the outer edges of the rings). The three sets of models on different orbits have very similar behavior in radial velocities at $R>6.5h$. However, there are some differences between the orbits V17 and V27 for the models with large $B/D$ ($B/D \ge 0.8$). In particular, there are secondary peaks at around $7h$ on the orbit V17 instead of on the orbit V27. This might happen because the initial relative velocity on V17 is lower and the integrating time of gravitational interactions is longer. Hence, it is possible to generate a secondary velocity structure at $7h$. Comparing to the clear relation between outer radial velocity profiles and $B/D$, the secondary velocity peaks are finer structures. Therefore, the increasing radial velocity outside the rings is more sensitive to $B/D$ of the progenitors than the initial relative velocity of the colliding pairs. The reason for this behavior is that the disk particles in a target galaxy are rotation-supported and are dynamically colder than the pressure-supported particles from the bulge. The bulge particles contain more components of radial motions than the disk particles. When the disks and the bulges are heated after collisions, the bulge particles expand more quickly than the disk particles. The expanding bulge particles with larger radial velocities result in the positive values of $v_{\rm R}(R)$ outside the ring. To confirm that the bulge particles contribute higher radial velocities in large radii, one needs to examine the mass fractions between the bulge and the stellar components and the radial velocities of the bulge particles. 

The outer-bulge-to-outer-stellar-mass (hereafter OBOS-mass) fractions of the progenitors and of the CRGs, at the radii where $R>5h$ and $R>6.5h$, are presented in Fig. \ref{bulgefrac}. The OBOS-mass fraction at $R>5h$ is defined as the mass ratio between the bulge and the stellar components in radii where $R>5h$ for a galaxy model. We remind the readers that $B/D$ is the overall mass ratio between the bulge and the disk for a galaxy. For CRGs, in the radii of $R > 5h$, the OBOS-mass fractions are smaller than $30\%$ even for the bulge-dominated models ($B/D=1.0$). At the position of the rings ($R=5h$), the OBOS-mass fractions are about $5\%-10\%$ smaller than those for the progenitors. However, in the radii where $R>6.5h$, the OBOS-mass fractions are significantly larger than those for the progenitors. For CRG models with $B/D \ge 0.3$, the OBOS-mass fractions are about $10\%-20\%$ larger than that for the progenitors. Moreover, the OBOS-mass fraction is larger than $50\%$ for CRGs with $B/D>0.4$ on the orbit of V17, and those for CRGs with $B/D>0.6$ on the orbits of V22 and V27. Consequently, the bulge mass dominates the stellar mass outside of the outer edge of the ring ($R>6.5h$). The OBOS-mass fraction of CRGs increases with growing values of $B/D$.

We find that the radial velocities of the bulge particles are from $140$ to $160\kms$ at $6.5h$, while the $v_{\rm R}(R=6.5h)$ of the disk particles range from $-50\kms$ to $10\kms$. Although the radial velocities of the bulge particles drop as the radius increases, the bulge particles are moving outwards outside of the outer edges of the rings (at $R>6.5h$). At the same time, the disk particles at $R> 6.5h$ are moving inwards. The very different behaviors for the $v_{\rm R}(R>6.5h)$ profiles of the two components imply that the motions of disk particles and the bulge particles are strongly decoupled. Therefore, the larger radial velocities of the overall stellar components at $R>6.5h$ for CRGs with larger $B/D$ come from the radial velocities of the bulge particles.  

In addition, in the lower limit of $B/D$ (with $B/D=0.0$), the $v_{\rm R}(R)$ profiles at large radii where $R> 6h$ turn into negative values. Therefore, for a pure-disk galaxy, the stellar particles beyond the radii of the rings are moving inwards. The ring forms through both the expansion of inner disk particles and the collapsing of the outer disk particles. 

In a CRG of small radii (i.e., within about $1h$), there is another positive velocity peak with a small amplitude of a few tens of $\kms$. Within the radii between $1h$ and the inner edge of the ring, a negative peak appears in the $v_{\rm R}(R)$ profile with an amplitude of $100-140~\kms$. These features of $v_{\rm R}(R)$ profiles indicate that the stellar particles within the inner edge of the ring are moving inwards the galactic center and that a secondary ring is forming. This issue has been extensively studied by earlier works on CRGs \citep{appleton1996collisional, gerber1996stellar, mapelli2012ring}. A ring is also composed of stellar particles from the outer particles collapsing inwards, in addition to the expansion of inner particles.

\subsubsection{Azimuthal Velocities}
\begin{figure*}
\centering
\includegraphics[width=180mm]{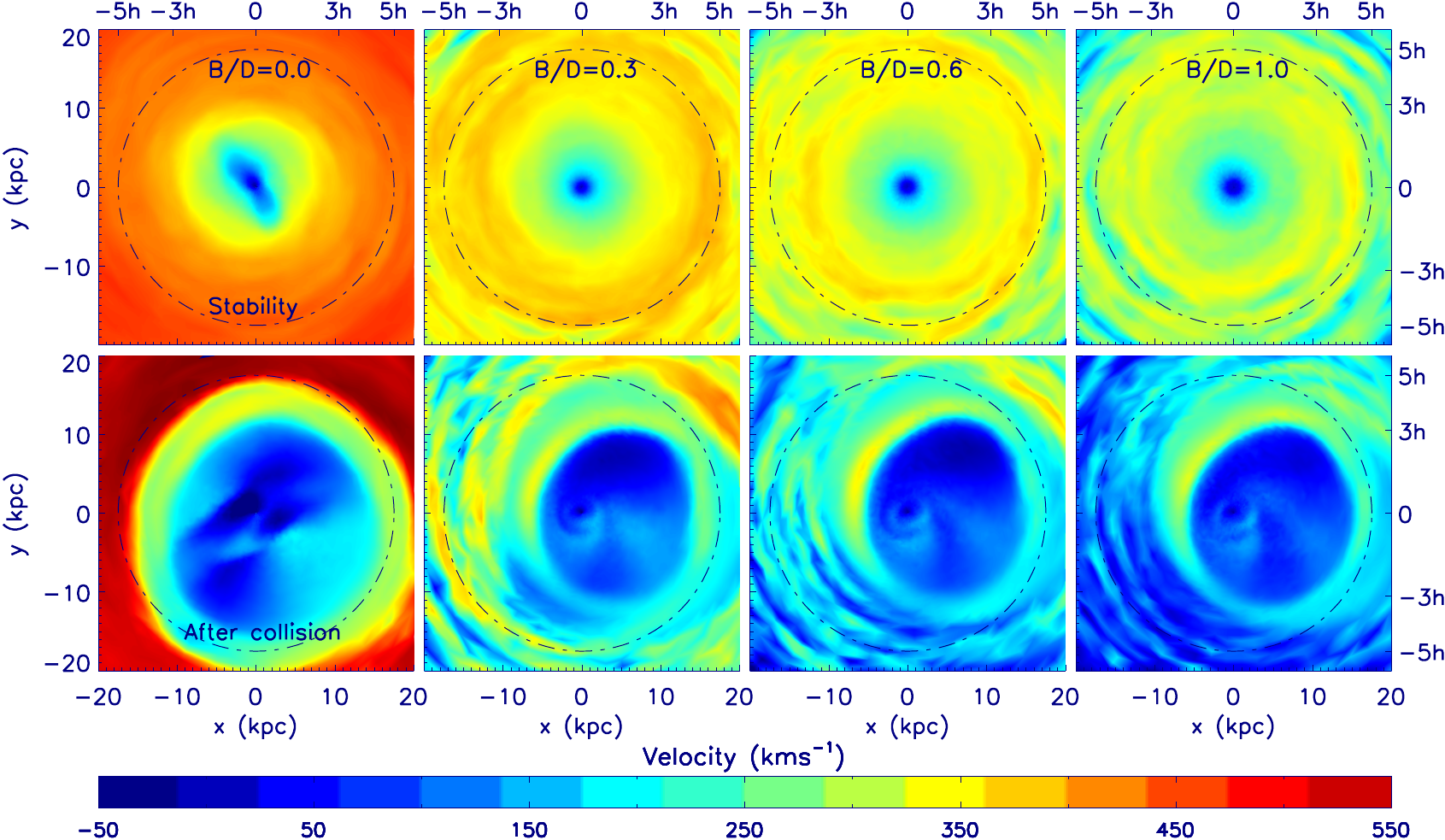}
\caption{The projected azimuthal velocity for models with different values of $B/D$ on the orbit of V22, when the rings propagate out to $x\approx 5h$ (dashed circles). The upper panels illustrate the progenitors and the lower panels illustrate the ring galaxies after collisions.}
\label{vphi_contour}
\end{figure*}	

The $v_{\rm \phi}(R)$ profiles for the same models and orbits are shown in the middle panels of Fig. \ref{BD_vr_vphi} with solid curves. The mass of a disk is smaller for larger value of $B/D$. Given that the rotation of the galaxy mainly comes from the disk particles, the average azimuthal velocity is larger for models with smaller values of $B/D$, especially in large radii. This relation still exists after the collisions. The main contribution to the azimuthal velocity comes from the disk particles. However, after the collisions, the $v_{\rm \phi}(R)$ profiles decrease within the radius between $1h$ and the inner edge of the ring. There is a velocity peak for the $v_{\rm \phi}(R)$ profile of a CRG at the radii of around $5h$. The $v_{\rm \phi}(R)$ profile declines more slowly in large radii as $B/D$ is lowered. When $B/D=0.3$, the $v_{\rm \phi}(R)$ profile is flat from the radius of the peak. For a model with $B/D<0.3$, the $v_{\rm \phi}(R)$ profile increases with the growth of radius in large radii. Thus, the azimuthal velocities of the stars keep increasing in the outer region of the rings for the disk-dominated galaxies. When $B/D=0.0$, the azimuthal velocity profiles are flat and the values are larger than the circular velocity (dashed curves, see Sec. \ref{vcirc}) in large radii where $R>6.5h$ for all the three models on different orbits. The overall azimuthal velocity profiles, $v_{\rm \phi}(R)$, which is combined with the azimuthal velocities from the disk particles and the bulge particles, are very different in large radii for CRGs with different $B/D$ values because the OBOS-mass fractions are different and the bulge particles almost do not contribute to the azimuthal velocities.

By combining the $v_{\rm R}(R)$ and $v_{\rm \phi}(R)$ profiles of the galaxies, we find that for a bulge-less or a disk-dominated galaxy, the stellar particles outside the ring structures rotate more rapidly and expand more slowly compared to that of the bulge-dominated galaxies.

In addition, for a given $B/D$, the shapes of $v_{\rm R}(R)$ and $v_{\rm \phi}(R)$ profiles do not change significantly with the growth of relative velocities for the galaxy pairs (on orbits V17, V22 and V27). Nevertheless, the $v_{\rm R}(R)$ and $v_{\rm \phi}(R)$ profiles are strongly influenced by the variation of $B/D$, while they are insensitive to the relative velocities.

We further study the projected azimuthal velocity maps on the disk plane for models on V22 orbit and show the results in Fig. \ref{vphi_contour}. There are four models, with $B/D=0.0,~0.3,~0.6$ and $1.0$. The upper panels present the $v_{\rm \phi}$ contours for the progenitors, and the lower panels show the $v_{\rm \phi}$ contours at $t_{\rm ring}$. After the collisions, $v_{\rm \phi}$ becomes smaller within the radius of the ring. Moreover, a clear, lopsided ring structure appears on the $v_{\rm \phi}$ maps after collision for all the models, which agrees with the azimuthal velocity peaks in the central panel of Fig. \ref{BD_vr_vphi}. For the model with $B/D=0.0$, the azimuthal velocity is even higher outside the position of the ring. Consequently, if the progenitor is a bulge-less galaxy, then the azimuthal velocity further grows as radius increases in the outer region of a ring galaxy. This stems from the fact that the progenitor is purely supported by rotation. The heating from the collision only produces a clean wave that propagates along the radial direction, without any extended envelopes. As $B/D$ increases, $v_{\rm \phi}$ is reduced outside the rings. In a bulge-dominated galaxy model (i.e., model with $B/D=1.0$), the azimuthal velocity is lower in the region inside the inner edge and outside of the outer edge of the ring after collision. In such a model, the collision mainly enhances the random motion for the stellar particles, especially for the bulge particles. 
\subsubsection{Vertical Velocities}
The vertical velocities, $v_{\rm z}(R)$, of the CRGs are shown in the lower panels of Fig. \ref{BD_vr_vphi}. 
There are clear peaks at the radius of the rings with amplitudes of $50$-$120\kms$. These peaks move with the propagation of the rings. The vertical velocities are higher for models with larger values of $B/D$ in large radii where $r \geq 5h$. This is due to the fact that the particles are perturbed through the collision of galaxies, not only in the radial directions but also in the vertical directions. This is consistent with the peaks and the outer extended structures for the $v_{\rm R}(R)$ profiles.

\subsection{Rotation Curves of CRGs}\label{vcirc}

The circular velocities on the disk planes, 
\beq v_{\rm circ}(R)\equiv \sqrt{R\frac{\partial \Phi(R)}{\partial R}},\eeq
are calculated numerically, where $\Phi(R)$ is the gravitational potential at the radius $R$ on the disk plane. The $v_{\rm circ}(R)$ profiles for CRGs with different initial values of $B/D$ formed on different orbits are shown (dashed curves) in the middle panels of Fig. \ref{BD_vr_vphi}. For a given radius in a CRG model, the value of $v_{\rm circ}(R)$ is larger than that of $v_{\rm \phi}$. There are two reasons for this. Firstly, the CRGs are not in equilibrium. The particles within $[1h,4h]$ are collapsing inwards and the particles forming the ring are expanding outwards. Secondly, the equilibrium systems with a bulge component are supported by both rotation and pressure. The rotation measured from the azimuthal velocities are smaller than the circular velocities because the velocity dispersion also supports the system in equilibrium. After the collisions, the bulge particles are heated through dynamical friction and, therefore, the velocity dispersion of the CRGs becomes much larger than that of the progenitors. This is the main reason. For the bulge-less models, the values of the $v_{\rm circ}(R)$ are slightly larger than that of $v_{\rm \phi}(R)$ in large radii because there are no bulge particles heated in these models.
	
\subsection{AGN-host Ring Galaxy SDSS J1634+2049}\label{SDSS,LWJ}
The ring in the host galaxy of the AGN, the SDSS J1634+2049 \citep{liu2016sdss}, is knotty and lopsided. Moreover, there are two companion galaxies with close values of redshifts on the projected sky. These are typical features of CRGs and we have demonstrated that the observed ring galaxy can be reproduced by modeling the off-center collisions between a disk galaxy and a dwarf galaxy. Note that in \citet{liu2016sdss}, the central region of the host galaxy is removed by using the {\it GALFIT} model \citep{haussler2007gems} and the ring structure behind the central region might also be removed. This explains why there is only an arc in the direction ahead of the central remnant.

The best fit of S\'{e}rsic index for the central remnant of the host galaxy of SDSS J1634+2049 obtained by \citet{liu2016sdss} is $n=4$, which implies that the central remnant of SDSS J1634+2049 is an early-type object. According to the analysis of Section \ref{sec-index_n}, the value of $B/D$ for the progenitor of this ring galaxy is in the range of $[0.1,~0.3]$. Therefore, the progenitor is a disk galaxy with a minor bulge component. For a pure-disk progenitor, the central remnant would appear to be a late-type object with $n<4$, while for a progenitor with $B/D>0.3$, the central remnant would contain a much more extended structure with $n>4$. 
	
Since the position for the ring and the separation for host galaxy of the SDSS J1634+2049 and C1 are very close to that of our generic CRG models, the velocity profiles, $v_{\rm R}(R)$, $v_{\rm z}(R)$ and $v_{\rm \phi}(R)$, for the host galaxy of the SDSS J1634+2049, should be very similar to that of the models with $B/D\in [0.1,~0.3]$ predicted in Fig. \ref{BD_vr_vphi}. As previously mentioned, the shapes of the velocity profiles are insensitive to the initial relative velocities of the galaxy pair but are sensitive to the values of $B/D$. We expect that future observations on the kinematics of the host galaxy of the SDSS J1634+2049 could provide further constraints to the structure of the progenitor. 

\subsection{CRGs with a Fixed Disk Mass}\label{fixdisk}

\begin{figure*}
\centering
\includegraphics[width=140mm]{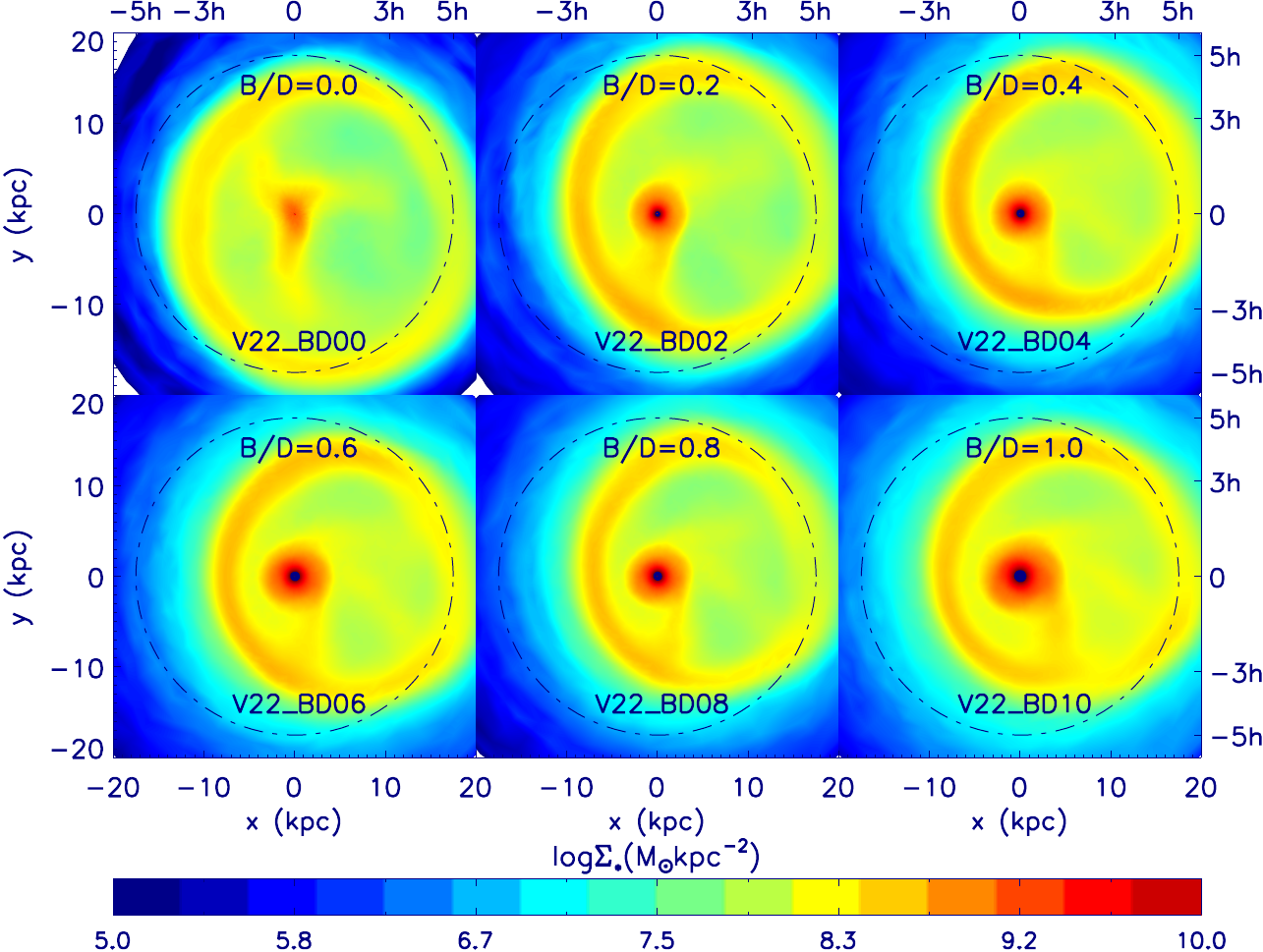}
\caption{The contours of the projected stellar-mass density of the ring galaxies models with a maintaining disk mass of $M_{\rm d}=1.03\times 10^{11}\msun$ and $B/D$ rising from $0.0$ to $1.0$ with an interval of $0.2$. The ring structure propagates out to $x \approx 5h$ (dashed circle). The colliding galaxy pair is on the orbit of V22.}
\label{fixedmd}
\end{figure*}

\begin{figure*}  
\centering  
\includegraphics[width=140mm]{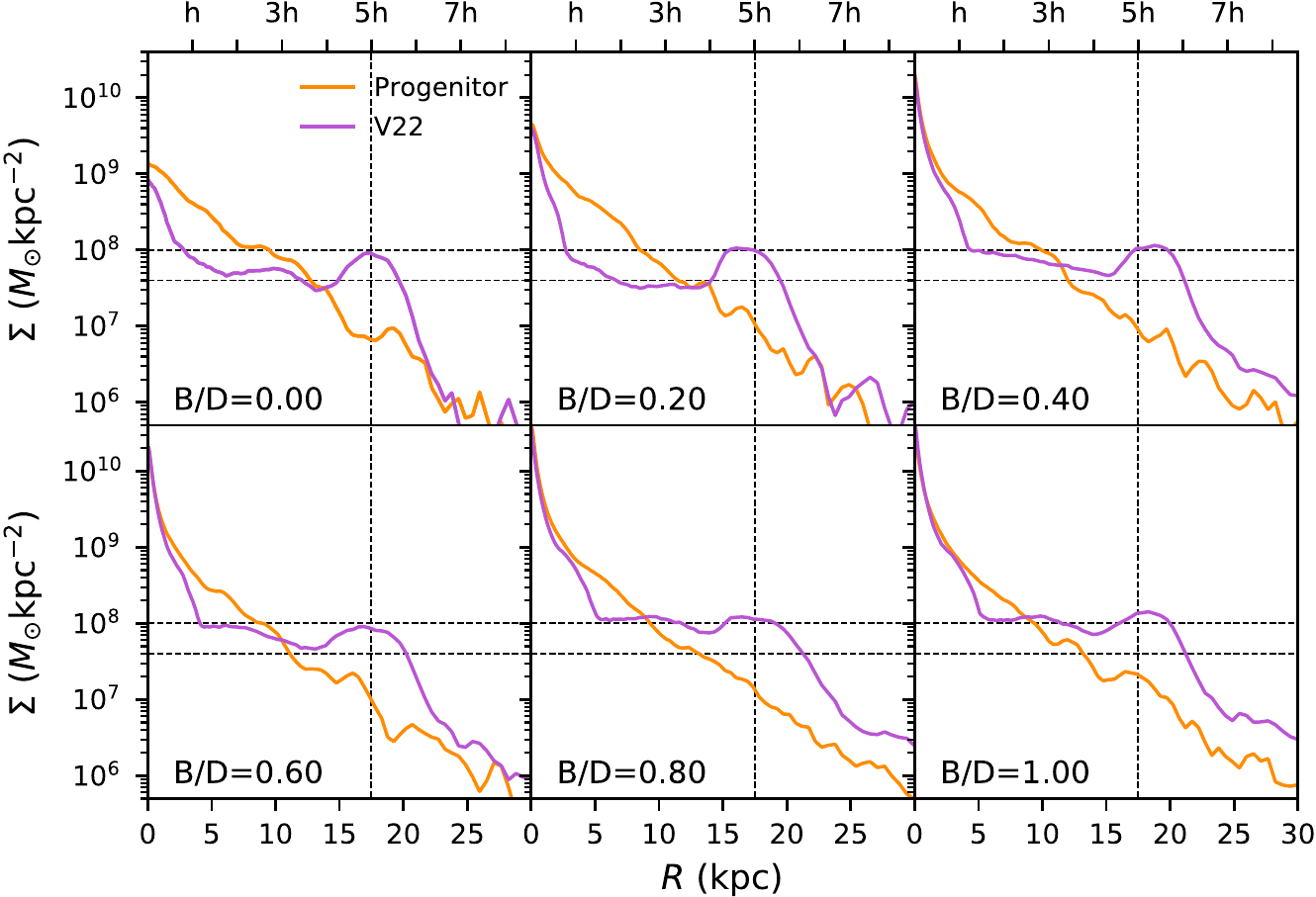}
\caption{Projected stellar-mass density profiles of the CRG with a fixed disk mass. The $B/D$ is ranging in $[0.0,~1.0]$ with an increasing interval of $0.2$. The colliding galaxy pair is on the orbit of V22.
}
\label{projden_fixedmd}
\end{figure*}

The morphologies and the projected density profiles of the CRGs with a fixed mass of disk are studied here. These CRG models are formed on the orbit of V22. The disk masses for the target galaxies are $M_{\rm d}=1.03\times 10^{11}\msun$. The values of $B/D$ increase from 0.0 to 1.0. The projected density contours for the CRGs are shown in Fig. \ref{fixedmd}, and the projected densities along x-axis are presented in Fig. \ref{projden_fixedmd}. The ring structures are the clearest when $B/D\in [0.0,~0.4]$, and they become more distinct with increasing $B/D$. Consequently, the trend for the disk-dominated galaxies to generate clearer ring structures in Fig. \ref{contourBD47} is not due to the larger mass of disks in those fixed-overall-stellar-mass models---the clearer ring structure is really related to smaller $B/D$.

\section{Conclusion and discussions}\label{conclusion}
We have revisited the impact of basic parameters on the formation of CRGs, including initial relative velocity, $\vzero$, the impact factor, $b$, the inclination angle, $\theta$, and the total mass and the scale length of the intruder galaxy. The results of the simulations agree with that of the existing studies \citep[e.g.,][and etc.]{lynds1976interpretation,theys1976ring,appleton1987models,appleton1996collisional,mapelli2012ring,smith2012numerical}. We have provided the details of these simulations in the Appendix \ref{appendix-parameters}. 
Moreover, we have found that the timescale for the rings spreading to the edge of the disk, $t_{\rm exist}$, correlates with $v_{\rm z0}$, the impact velocity's component vertical to the plane of the target galaxy. This indicates that in a higher speed encounter, the propagation velocity of a ring slows down in the later stage and the ring can exist on the disk plane for a longer timescale. This is because the integration time of gravitational interaction is shorter in a higher speed encounter, and the integrated perturbation from the intruder galaxy is weaker. 

We have studied the influence of different values of $B/D$ for the progenitor on the CRG formation in this work. Three orbits are selected from the examinations of basic parameters. These orbits can successfully reproduce a clear ring structure at $\approx 5h$ when the intruder galaxy is $10h\pm 2h$ away projected on the disk plane of the target galaxy. The values of $B/D$ for the target galaxy varies in the range $[0.0,~1.0]$ with an increasing interval of $0.1$. Bulges with lower mass lead to clearer ring structure. The projected density profiles of the CRGs have further shown that the ring is the strongest with a pure-disk or a disk-dominated progenitor.

The S\'{e}rsic indices for the central remnants of the CRGs have also been studied in this work. A strong correlation between $B/D$ and $n$ has been found. A bulge-less target galaxy leaves a late-type central remnant in the CRG, while the progenitors with a bulge component produce early-type central remnants. Therefore, for an observed CRG, the $B/D$ value of the progenitor can be predicted by studying the S\'{e}rsic index for the central region in the ring galaxy.

Moreover, we have found that the shapes of the velocity profiles, $v_{\rm R}(R)$ and $v_{\rm \phi}(R)$, depend sensitively on the concrete value of $B/D$. A bulge-less or a disk-dominated target galaxy results in smaller values of $v_{\rm R}(R)$ and larger values of $v_{\rm \phi}(R)$ in large radii; i.e., outside of the outer edge of the ring. A bulge-dominated galaxy leads to higher expanding velocity in the radial direction and more rapidly declining $v_{\rm \phi}(R)$ in large radii. The reason for this is that the heating from the collision enhances the radial motions of the bulge particles. Therefore, for a model with larger values of $B/D$, there are more radial motions in large radii after the collision. In contrast, the radial and azimuthal velocities in large radii are immune to the initial orbital velocities of the colliding galaxy pairs.
In addition, the rotation curves of the CRGs are studied. The rotation curves are above the $v_{\rm \phi}(R)$ curves for CRGs with $B/D>0.0$ because the CRGs are partially supported by velocity dispersion. Another reason is that the CRGs are not in an equilibrium state.

Finally, we have applied our results to a newly observed AGN-host ring galaxy, the SDSS J1634+2049. The ring morphology and the projected distance of the companion galaxy can be well-explained by our CRG models colliding on three orbits. We have further predicted that the progenitor of this ring galaxy is a late-type disk galaxy with a minor bulge component and $B/D \in [0.1,~0.3]$.

The separation between the SDSS J1634+2049 and its companion C1 galaxy is up to $35.7\kpc$ projected on sky. Since the propagation time for an outer ring is at a timescale of $100 \Myr$, the initial relative velocity, $v_0$, between the two galaxies might be too high to explain its origin. Suppose that the intruder galaxy is at infinity to the target galaxy at the beginning. Therefore, the two galaxies are approaching each other owing to the attraction caused by gravitation. The relative velocity of the two galaxies at a radius $r$ is the free falling velocity of the intruder galaxy in the gravitational potential well of the target galaxy; i.e, the escape velocity, $v_{\rm esc}$. The escape velocity satisfies $v_{\rm esc}^2/2+\Phi=0$. For simplicity, we take $v_{\rm esc}=\sqrt{2GM/r}$ for a point-mass galaxy with total mass $M$, and $v_{\rm esc}\approx 805\kms$ for the target galaxy model in this work. For a real galaxy with an extended density distribution, the actual escape velocity is smaller than that obtained under the point-mass approximation. However, the point-mass approximation can be used to make a quick estimation. We have found that $v_0=640\kms,~720\kms$ and $860\kms$ in the three possible orbits (i.e., orbits V17, V22 and V27) suggested by our simulations. The origin of $v_0$ between the galaxy pair on orbits V17 and V22 can be easily explained by the free falling of dwarf galaxy. However, $v_0$ between the galaxy pair on orbit V27 is slightly larger than $v_{\rm esc}$, which is in the critical limit. In summary, the values of $v_0$ introduced in our models are plausible for all the three orbits.

\section{Acknowledgements}
We thank the anonymous referee for careful and helpful suggestions and comments to the manuscript. The authors thank Ning Jiang for the discussion on the S\'{e}rsic model fitting for the observed AGN-host ring galaxy, the SDSS J1634+2049, and Yongda Zhu and Xiaobo Dong for the discussion about the usage of the \textit{N}-body code, {\it RAMSES}. An early stage of the numerical calculations in this paper have been done on the supercomputing system in the Supercomputing Center of University of Science and Technology of China. This work is supported by the National Natural Science Foundation of China (NSFC 11503025, 11421303), Anhui Natural Science Foundation under grant 1708085MA20 and ``the Fundamental Research Funds for the Central Universities''. X.W. thanks for the additional support provided by ``Hundred Talents Project of Anhui Province''. X.K. and G.C. are supported by the National Key R\&D Program of China (2015CB857004, 2017YFA0402600), and the National Natural Science Foundation of China (NSFC, Nos. 11320101002, 11421303, and 11433005). G.C. thanks for support by ``Fund for Fostering Talents in Basic Science of the National Natural Science Foundation of China NO.J1310021''. W.L. acknowledges supports from the Natural Science Foundation of China grant (NSFC 11703079) and the "Light of West China" Program of Chinese Academy of Sciences (CAS). 

\bibliographystyle{aasjournal}
\bibliography{CRG}

\begin{appendix}

\section{Details of Parameters for the Colliding Galaxy Pair}\label{appendix-parameters}

\subsection{Initial relative velocities}\label{v0}
We set up a series of initial relative velocities, which reproduces from low to high speed encounters. The parameters for the orbits of the colliding systems are listed in columns $2$--$6$ of Table \ref{orbitpara}. The initial distances of the two galaxies are in the range of $[37.3, ~102.8]\kpc$, standing for typical distances between a host galaxy and a satellite galaxy. The location of the intruder galaxy is within the virial radius of the DM halo of the disk galaxy. The morphology of the disk galaxy will be studied after collisions and the most suitable orbits which produce ring structures will be selected.

\begin{table*}
\centering
\caption{Parameters of the Collisional Galaxy Pair}\vskip 0.5cm
\begin{threeparttable}
\begin{tabular}{lcccccccccccccccccccc}
\hline
\hline
Collisional & $z_0$ & $b$ &$v_{o}$ & $v_{xo}$ & $v_{zo}$ & $\theta$ & $t_{\rm{gen}}$ & $t_{\rm{ring}}$ & $t_{\rm{exist}}$ &$M_{\rm{ring}}$ \\
models & (kpc) & (kpc) & ($\kms$) & ($\kms$) & ($\kms$) & ($^\circ$) & (Myr) & (Myr) & (Myr) & ($10^{10}\Msun$)\\
\hline
V01 & -16.1 & 1.6 & 447   & 400 & 200 & 63.4 & 17.6 & 91.3  & 145.5 & 1.90 \\
V02 & -10.7 & 1.1 & 632   & 600 & 200 & 71.6 & 16.9 & 109.2 & 135.5 & 1.06 \\
V03 & -16.1 & 1.6 & 559   & 500 & 250 & 63.4 & 20.0 & 100.8 & 143.3 & 1.63 \\
V04 & -32.1 & 2.5 & 424   & 300 & 300 & 45.0 & 19.4 & 70.8  & 150.8 & 4.69 \\
V05 & -24.1 & 2.1 & 500   & 400 & 300 & 53.1 & 21.4 & 87.6  & 137.0 & 3.38 \\
V06 & -19.3 & 1.8 & 583   & 500 & 300 & 59.0 & 19.5 & 99.2  & 150.7 & 2.09 \\
V07 & -16.1 & 1.6 & 671   & 600 & 300 & 63.4 & 16.6 & 91.0  & 159.4 & 3.16 \\
V08 & -16.1 & 1.6 & 783   & 700 & 350 & 63.4 & 17.7 & 87.7  & 147.6 & 3.49 \\
V09 & -64.3 & 3.2 & 447   & 200 & 400 & 26.6 & 14.4 & 54.8  & 174.4 & 3.16 \\
V10 & -42.8 & 2.9 & 500   & 300 & 400 & 36.9 & 16.9 & 66.1  & 170.1 & 4.01 \\
V11 & -25.7 & 2.2 & 640   & 500 & 400 & 51.3 & 24.7 & 73.0  & 165.8 & 4.57 \\
V12 & -21.4 & 2.0 & 721   & 600 & 400 & 56.3 & 21.5 & 81.1  & 169.7 & 4.29 \\
V13 & -16.1 & 1.6 & 894   & 800 & 400 & 63.4 & 18.8 & 84.2  & 164.8 & 3.67 \\
V14 & -16.1 & 1.6 & 1,006 & 900 & 450 & 63.4 & 19.7 & 86.8  & 172.4 & 3.38 \\
V15 & -64.3 & 3.2 & 559   & 250 & 500 & 26.6 & 14.4 & 46.2  & 185.6 & 2.63 \\
V16 & -53.6 & 3.1 & 583   & 300 & 500 & 31.0 & 18.4 & 58.6  & 175.8 & 3.39 \\
V17 & -40.2 & 2.8 & 640   & 400 & 500 & 38.7 & 19.9 & 60.1  & 167.1 & 3.45 \\
V18 & -26.8 & 2.3 & 781   & 600 & 500 & 50.2 & 20.3 & 71.2  & 175.7 & 4.07 \\
V19 & -23.0 & 2.1 & 860   & 700 & 500 & 54.5 & 22.8 & 71.5  & 175.7 & 3.95 \\
V20 & -96.4 & 3.4 & 632   & 200 & 600 & 18.4 & 17.4 & 46.9  & 183.3 & 2.82 \\
V21 & -64.3 & 3.2 & 671   & 300 & 600 & 26.6 & 13.6 & 50.6  & 179.2 & 3.07 \\
V22 & -48.2 & 3.0 & 721   & 400 & 600 & 33.7 & 18.3 & 56.6  & 179.1 & 3.24 \\
V23 & -38.6 & 2.7 & 781   & 500 & 600 & 39.8 & 18.9 & 56.6  & 182.6 & 3.07 \\
V24 & -32.1 & 2.5 & 849   & 600 & 600 & 45.0 & 19.5 & 62.0  & 181.7 & 3.34 \\
V25 & -21.4 & 2.0 & 1,082 & 900 & 600 & 56.3 & 19.8 & 79.4  & 178.6 & 3.32 \\
V26 & -64.3 & 3.2 & 783   & 350 & 700 & 26.6 & 17.2 & 46.4  & 184.3 & 2.69 \\
V27 & -45.0 & 2.9 & 860   & 500 & 700 & 35.5 & 19.2 & 55.9  & 181.4 & 3.04 \\
V28 & -48.2 & 3.0 & 1,082 & 600 & 900 & 33.7 & 18.1 & 51.0  & 187.8 & 2.62 \\
V29 & -32.1 & 2.5 & 1,273 & 900 & 900 & 45.0 & 19.0 & 65.1  & 201.4 & 2.83 \\
\hline
\end{tabular}
 \begin{tablenotes}
\item 
\begin{flushleft}
\textbf{Note.} The initial positions of the intruder galaxy are $(-10.0h,~0.0, ~z_0)$ and the values of $z_0$ are listed in column 2. The impact parameter $(b)$, the initial relative velocity $(\vzero )$ and the components of the initial relative velocity ($v_{\rm x0}$ and $v_{\rm z0}$) are shown in columns 3--6. $\theta$ is the inclination angle for the collisional galaxy pair. $t_{\rm{gen}}$ is the timescale between the collision and the formation of the ring. $t_{\rm{ring}}$ is the timescale that a ring propagates out to $ 5h\approx 18\kpc$ after collision, and $t_{\rm{exist}}$ is the timescale of the ring spreading to the edge of the galaxy. $M_{\rm{ring}}$ is the mass of the ring when it propagates out to $5h$.
\end{flushleft}
\end{tablenotes}
 \label{orbitpara}
\end{threeparttable}
\end{table*} 	

\begin{figure}
\centering
\includegraphics[width=90mm]{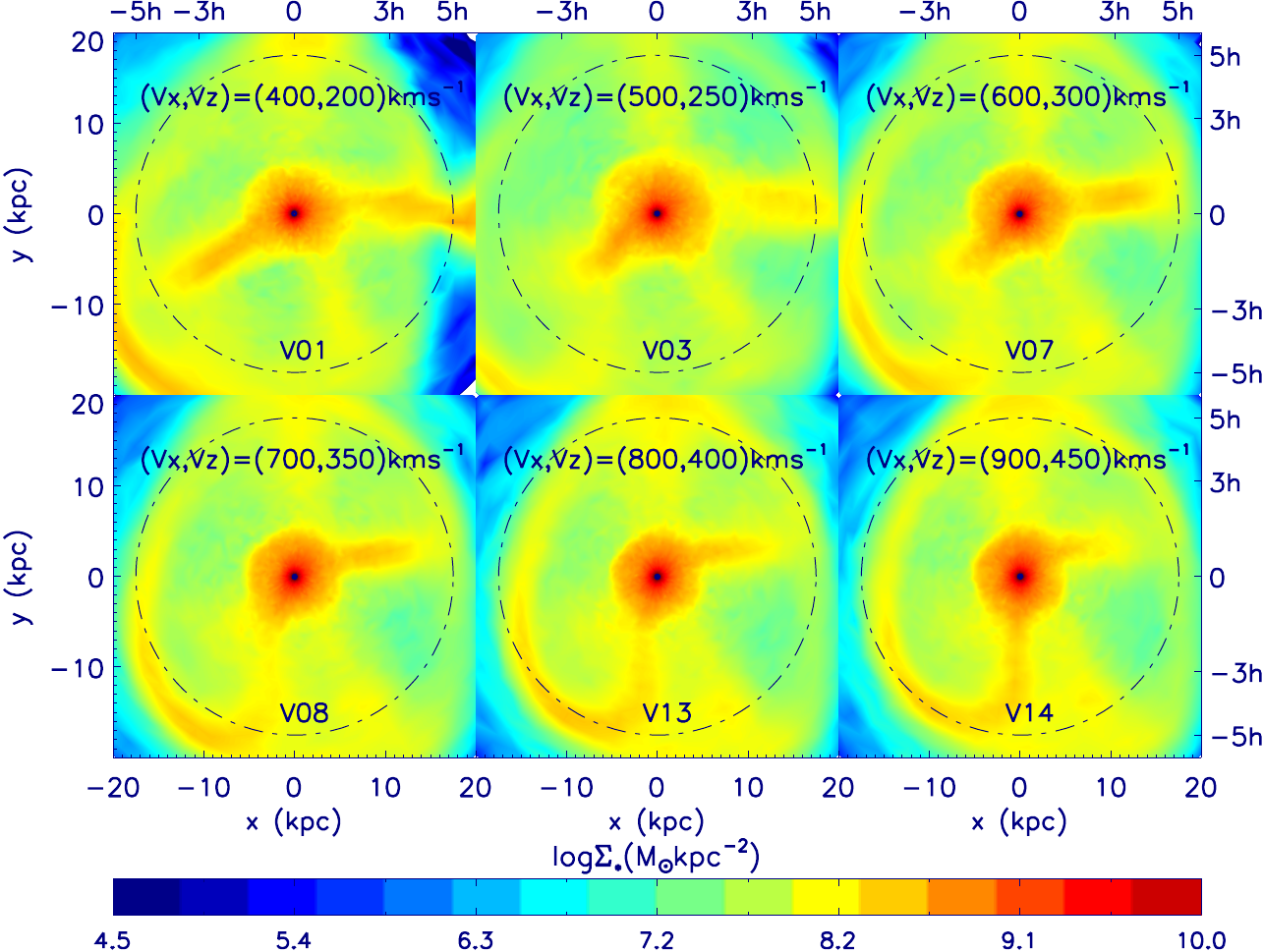}\vskip 0.50cm
\caption{The contours of the projected stellar-mass density of the target galaxy when the ring structure propagates out to $\approx 5h$ along positive $x$-axis. The models have the same values of $z_0$, $b$ and $\theta$, but different values of $\vzero$.}
\label{vcontour1}
\end{figure}

As previously mentioned, by testing a number of different values of $\vzero$, the orbital parameters of the intruder galaxy are determined and summarized in Table \ref{orbitpara}. A series of \textit{N}-body simulations are implemented for a few hundred $\Myr$ till the intruder galaxy passes though the disk galaxy and a ring structure propagates out to $\approx 5h\sim 18\kpc$, and the propagation timescale, $t_{\rm ring}$, is listed in column 9 of Table \ref{orbitpara}. Fig. \ref{vcontour1} shows six model CRGs (V01, V03, V07, V08, V13 and V14). We select the models moving along the same direction but with different values of $\vzero$. According to Equation. \ref{z0b}, the values of $z_0$, $b$ and $\theta$ are the same for the six models. With the increase of $\vzero$, the propagation timescale for the ring is shorter and the ring becomes less clear. This happens because the integration time of interaction from the companion galaxy is shorter for higher speed encounters. In addition, the rings are asymmetric and incomplete. These results agree with the existing studies \citep[e.g.,][]{mapelli2008ring,smith2012numerical}. 

\subsection{Inclination Angles}
\begin{figure*}
\centering
\includegraphics[width=180mm]{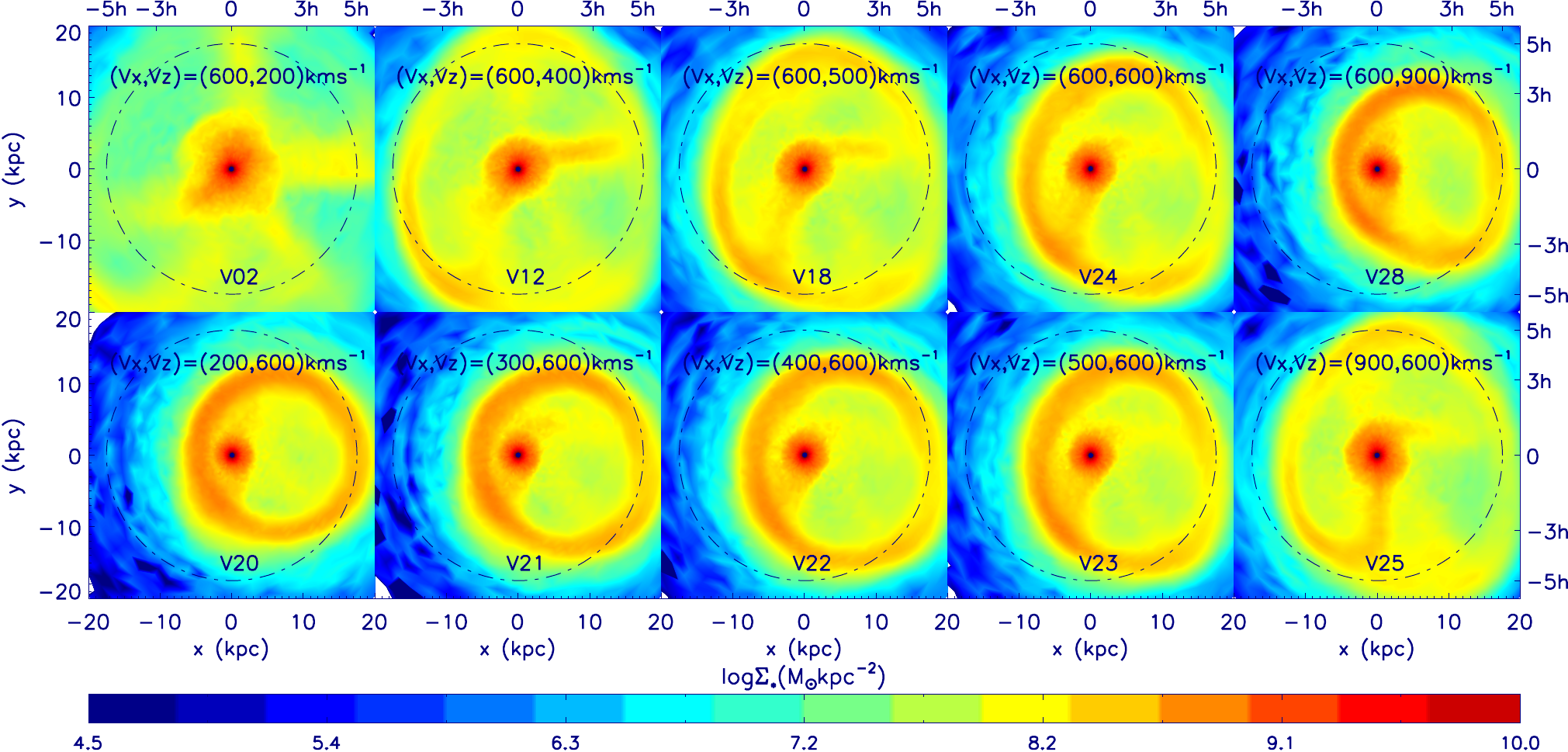}
\caption{The contours of the projected stellar-mass density for the ring galaxy when the ring propagates out to $\approx 5h$ along positive $x$-axis. The models are selected by different values of $\theta$, which is controlled through a fixed value of $v_{\rm x0}=600~\kms$ and varying $v_{\rm z0}$ (upper panels), or through a fixed value of $v_{\rm z0}=600~\kms$ and varying $v_{\rm x0}$.}
\label{vcontour2}
\end{figure*}

Two sets of models in Table \ref{orbitpara} are selected to test the effect of inclination angles, $\theta$, in CRGs. The models are selected by a fixed value of $v_{\rm x0}=600~\kms$ (galaxy-pair models V02, V12, V18, V24, V28), or by a fixed value of $v_{\rm z0}=600~\kms$ (galaxy-pair models V20, V21, V22, V23, V25), corresponding to galaxy pairs with descending and ascending values of $\theta$, respectively. The projected stellar density contours for the ring galaxies are presented in Fig. \ref{vcontour2}. 

The completeness of the ring increases as $\theta$ decreases. Although the initial relative velocity increases significantly with decreasing $\theta$ for the models in upper panels of Fig. \ref{vcontour2}, the ring structure is becoming more clear. This implies that the density of the ring is higher with smaller $\theta$, and that $\theta$ influences the ring morphology more strongly than the initial relative velocity does. Note that the ring structure does not propagate out to the radius of $5h$ for model V28 because the initial relative velocity is too high and the integrating time for gravitational interactions between the two galaxies is not long enough. 

In the other set of models (i.e., models in the lower panel of Fig. \ref{vcontour2}) the ring structure becomes more blurred with increasing values of both $\theta$ and initial relative velocity. For the model with largest values of $\theta$ and initial relative velocity (i.e., model V25), the ring nearly vanishes when it spreads out to the radius of $5h$ (e.g., \citealt{fiacconi2012adaptive, wu2015formation}).

\subsection{Impact Factor}\label{impact}

\begin{table}
\centering
\caption{Parameters of Galaxy-pair Models with Different Values of Impact Factor, $b$}\vskip 0.2cm
\begin{threeparttable}
\begin{tabular}{lcccccccccccccc}
\hline
\hline
Collisional & $z_0$& $b$ & $t_{\rm{gen}}$ & $t_{\rm{ring}}$ & $t_{\rm{exist}}$ &$M_{\rm{ring}}$ \\
models & (kpc) & (kpc) & (Myr) & (Myr) & (Myr) & ($10^{10}\Msun$)\\
\hline
b01 & -53.6 & 0.0 & 24.1 & 96.5 & 187.9 & 2.82 \\
b02 & -50.9 & 1.5 & 21.5 & 66.5 & 188.6 & 3.69 \\
b03 & -48.2 & 3.0 & 18.3 & 56.6 & 179.1 & 3.24 \\
b04 & -45.5 & 4.5 & 14.4 & 52.7 & 158.3 & 2.88 \\
b05 & -42.8 & 5.9 & 16.6 & 55.1 & 151.3 & 2.27 \\
b06 & -40.2 & 7.4 & 17.8 & 49.5 & 143.6 & 1.90 \\
\hline
\end{tabular}
\label{b,time}
 \begin{tablenotes}
\item 
\begin{flushleft}
\textbf{Note.} The initial relative velocity and inclination angle are the same as that of model V22 in Table \ref{orbitpara}. The initial position of the intruder galaxy is $(-10.0h,~0.0, ~z_0)$ and the values of $z_0$ are listed in  column 2. The impact parameter $(b)$, are shown in column 3.
\end{flushleft}
\end{tablenotes}
\end{threeparttable}
\end{table}

\begin{figure}
\centering
\includegraphics[width=90mm]{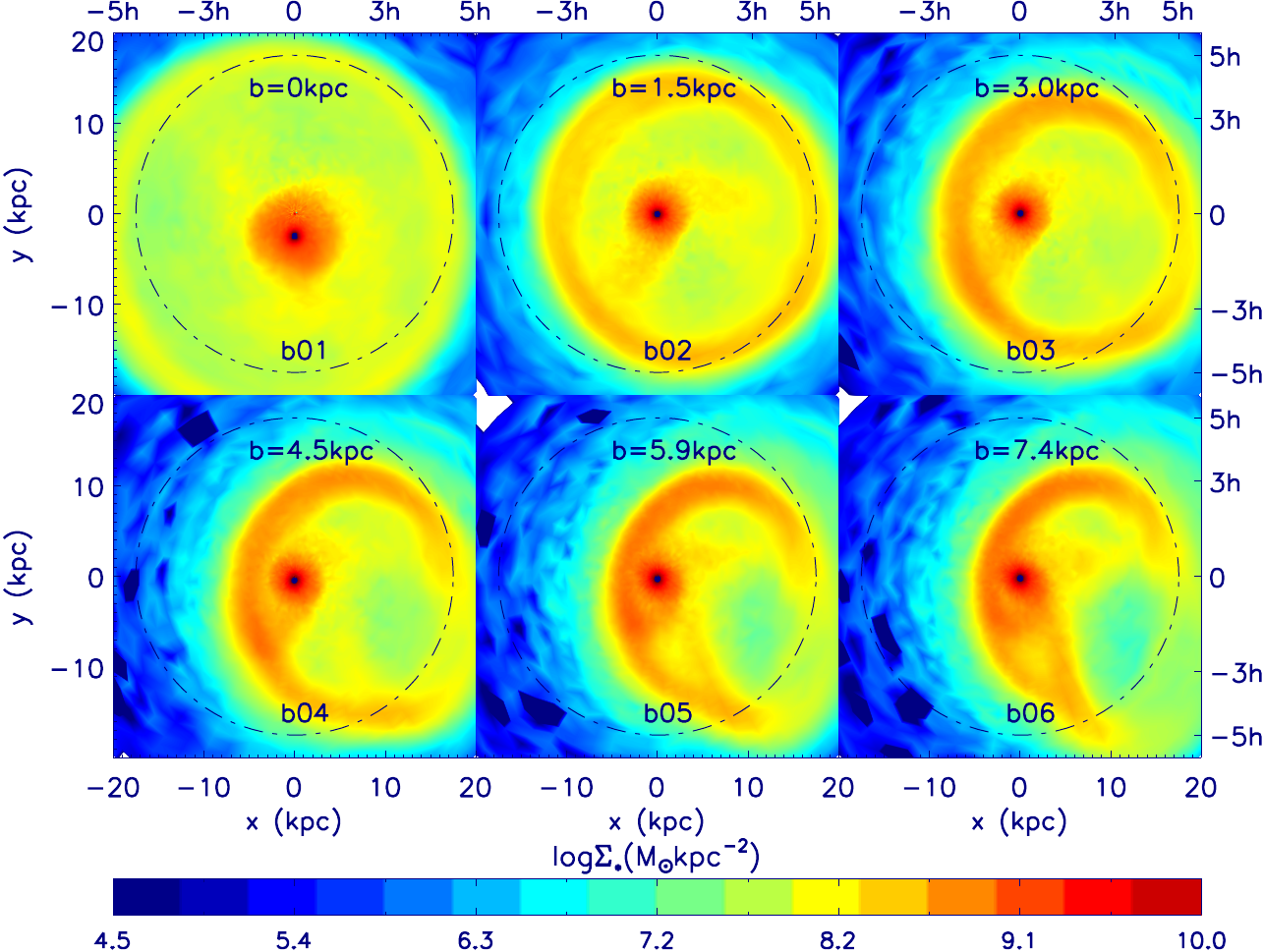}
\caption{The contour plots of the projected stellar-mass density of the target galaxy when the $+x$ ring structure propagates out to $\approx 5h$. $b$ grows up from the left-hand panel to the right-hand panel and all panels are with fixed ($v_{\rm x0}, v_{\rm z0})=(400, 600)~\kms$.}
\label{bcontour}
\end{figure}

The effect of impact parameter for a colliding galaxy-pair will be studied here. We select an orbit from the above simulations, namely: model V22. The initial relative velocity is ($v_{\rm x0}, v_{\rm z0})=(400, 600)~\kms$ and the inclination angle is $33.7\deg$. The initial relative velocity is below the escape velocity of the target galaxy at the radius of the companion galaxy and the inclination angle is moderate. The impact factor varies in the range of $[0.0,~7.4]\kpc = [0.0,~2.1h]$ and is listed in Table \ref{b,time}. Fig. \ref{bcontour} presents the projected density contours for CRGs with varying impact parameter of the colliding galaxy pairs. When the value of $b$ increases, the ring becomes more asymmetric. The half-mass radius of the progenitor is denoted as $R_{\rm e,0}$, which is approximately $5\kpc \approx 1.4h$ (see $L_{\rm 0.5}$ in the upper left-hand panel of Fig. \ref{lr}). When $b\gtrsim R_{\rm e,0}$ (i.e., models b05, b06 in Table \ref{b,time}), the density peaks (bulge) are almost embedded in the rings. Moreover, there appear some spiral arm-like structures connecting the rings and the bulges appear. These results agree well with the previous numerical simulations of RE galaxies by \cite{lynds1976interpretation}, \cite{gerber1992model}, and \cite{mapelli2012ring}. 

\subsection{Mass and Size of the Intruder Galaxy}

\begin{table}
\centering
\caption{Parameters of the Collisional Galaxy Pair, Model V22, with Different Values of $M_{\rm P}$ and $r_{\rm P}$} \vskip 0.2cm
\begin{threeparttable}
\begin{tabular}{lcccccccccccc}
\hline
Collisional & $M_{\rm{P}}$ & $r_{\rm P}$ & $t_{\rm{gen}}$ & $t_{\rm{ring}}$ & $t_{\rm{exist}}$ &$M_{\rm{ring}}$ \\
models & $10^{11}\Msun$ & (kpc) & (Myr) & (Myr) & (Myr) & ($10^{10}\Msun$)\\
\hline
I01 & 1.0 & 3.8 & 35.9 & 106.9 & 203.5 & 1.14 \\
I02 & 1.0 & 1.9 & 28.7 & 102.8 & 201.4 & 1.21 \\
I03 & 1.0 & 1.2 & 25.4 & 100.8 & 184.8 & 1.30 \\
I04 & 2.0 & 3.8 & 26.5 & 95.8  & 222.1 & 1.75 \\
I05 & 2.0 & 1.9 & 17.1 & 57.1  & 203.4 & 2.21 \\
I06 & 2.0 & 1.2 & 18.9 & 44.5  & 201.5 & 2.85 \\
I07 & 3.0 & 3.8 & 19.8 & 71.5  & 200.4 & 2.01 \\
I08 & 3.0 & 1.9 & 14.2 & 45.5  & 174.2 & 3.07 \\
I09 & 3.0 & 1.2 & 12.8 & 34.7  & 163.2 & 3.07 \\
\hline
\end{tabular}
\label{Intruder,time}
 \begin{tablenotes}
\item 
\begin{flushleft}
\textbf{Note.} The total mass and scale length of intruder galaxies are shown in columns 2 and 3. The initial position of intruder galaxies are $(-10.0h,~0.0, 18.4h)$, $b = 3.2\kpc\approx h$ and ($v_{\rm x0}, v_{\rm z0})=(300, 600)~\kms$. Model I05 is the model V21 in Table \ref{orbitpara}.
\end{flushleft}
\end{tablenotes}
\end{threeparttable}
\end{table}

\begin{figure}
\centering
\includegraphics[width=90mm]{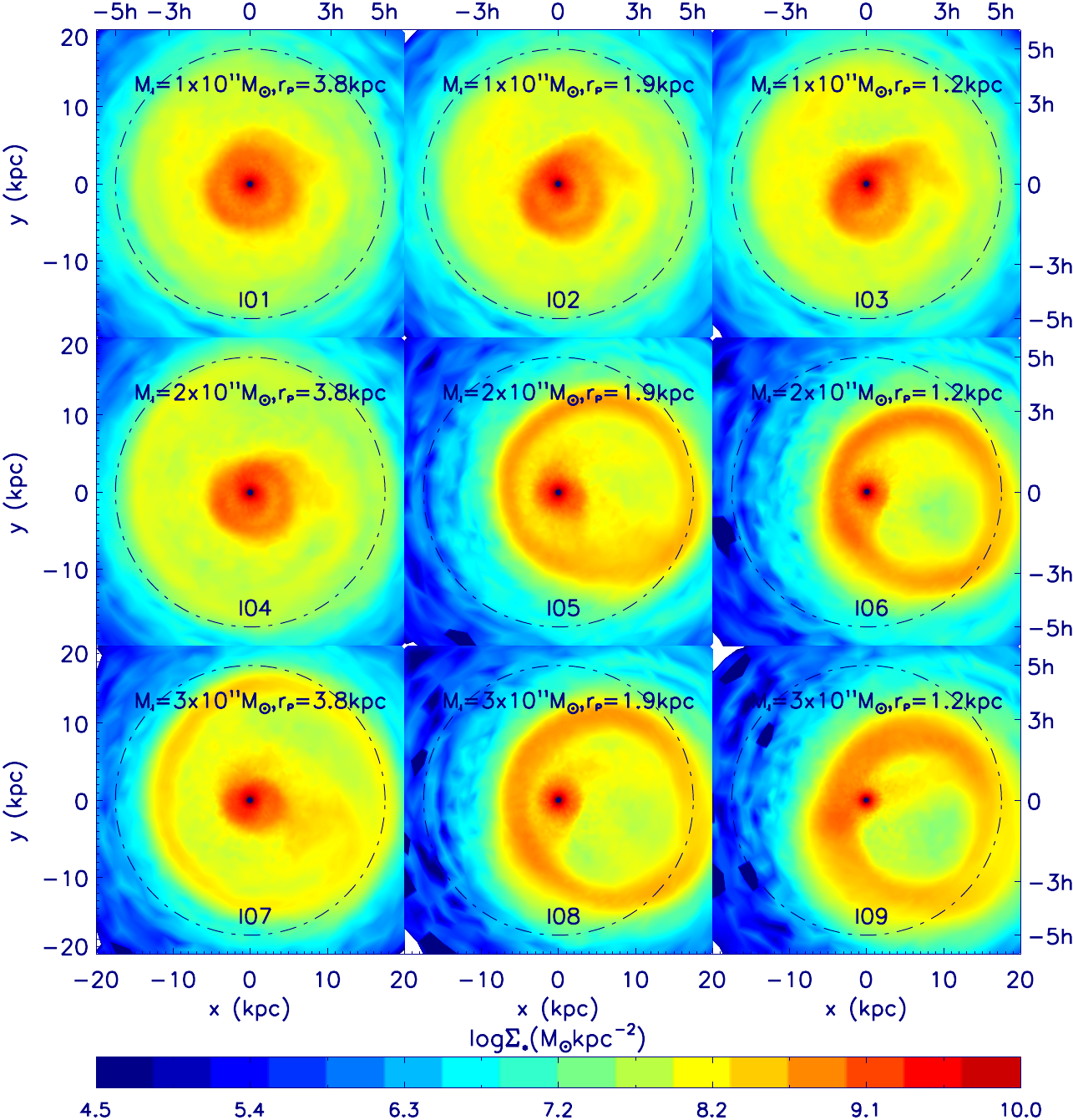}
\caption{The contours of the projected stellar-mass density of the target galaxy when the ring structure propagates out to $\approx 5h\sim 18\kpc$ along positive $x$-axis. The total mass of intruder galaxies increases from the top to the bottom panels and the scale length of intruder galaxies decreases from the left-hand to the right-hand panels.}
\label{Icontour}
\end{figure}

The effect of the different total mass and scale length of an intruder galaxy on an orbit of model V21 has been tested. The model parameters are listed in columns $2$ and $3$ of Table \ref{Intruder,time}. Fig. \ref{Icontour} shows the ring structures formed in the colliding galaxy pairs by tuning mass and size (i.e., half-mass radius) of the intruder galaxy. The mass of the intruder galaxy increases from $1\times 10^{11}\msun$ to $3\times 10^{11}\msun$ (the top panels to the bottom panels), and the half-mass radius decreases from $3.8\kpc \approx 1.1h$ to $1.2\kpc \approx 0.3h$ (from the left-hand panels to the right-hand panels). No rings form when the mass for the dwarf galaxy is $1\times 10^{11}\msun$, which is $\approx 3\%$ of the mass for the target galaxy.
The projected density of the ring is larger when the mass of the intruder galaxies increases, and the ring is clearer when the size (i.e., the half-mass radius) of the intruder galaxy decreases. Thus, a clearer galactic ring forms when there is a more compact and more massive intruder galaxy. It agrees with the previous modeling of CRGs by \citet{appleton1996collisional} and \citet{smith2012numerical}.

\end{appendix}
\end{document}